\documentclass[float,epsfig,usenatbib]{mn2e}

\usepackage{graphicx}
\usepackage{appendix}
\usepackage{psfrag}
\usepackage{amsmath,amssymb}
\usepackage{threeparttable}
\usepackage{float}
\usepackage{subfigure}
\usepackage{afterpage}

\title[The GALEX Arecibo SDSS Survey. IV.]{The GALEX Arecibo SDSS
  Survey. IV. Baryonic Mass-Velocity-Size Relations of Massive Galaxies}
\author[B. Catinella et al.]
{Barbara Catinella$^{1}$\thanks{bcatinel@mpa-garching.mpg.de},
  Guinevere Kauffmann$^{1}$, David Schiminovich$^{2}$, 
  \newauthor Jenna Lemonias$^{2}$, Cecilia Scannapieco$^{3}$, Jing Wang$^{1,4}$, Silvia Fabello$^{1}$, 
  \newauthor Cameron Hummels$^{2}$, Sean M. Moran$^{5}$, Ronin Wu$^{6,7}$, Andrew P. Cooper$^{1}$, 
  \newauthor Riccardo Giovanelli$^{8}$, Martha P. Haynes$^{8}$, Timothy M. Heckman$^{5}$,
  \newauthor and Am{\'e}lie Saintonge$^{9}$\\
$^{1}$Max-Planck Institut f\"{u}r Astrophysik, D-85741 Garching, Germany\\
$^{2}$Department of Astronomy, Columbia University, New York, NY 10027, USA\\
$^{3}$Leibniz-Institut f\"{u}r Astrophysik Potsdam (AIP), D-14482 Potsdam, Germany\\
$^{4}$Center for Astrophysics, University of Science and Technology of China, 230026 Hefei, China\\
$^{5}$Department of Physics and Astronomy, The Johns Hopkins University, Baltimore, MD 21218, USA\\
$^{6}$Commissariat \`a l'Energie Atomique (CEA), 91191 Gif-sur-Yvette, France\\
$^{7}$Department of Physics, New York University, New York, NY 10003 USA\\
$^{8}$Center for Radiophysics and Space Research, Cornell University, Ithaca, NY 14853, USA\\
$^{9}$Max Planck Institut f\"{u}r extraterrestrische Physik, D-85741 Garching, Germany\\
}

\date{}

\begin{document}

\def\deg{$^{\circ}$}
\newcommand{\eg}{e.g.}
\newcommand{\ie}{i.e.}
\newcommand{\minusone}{$^{-1}$}
\newcommand{\kms}{km~s$^{-1}$}
\newcommand{\kmsm}{km~s$^{-1}$~Mpc$^{-1}$}
\newcommand{\Ha}{$\rm H\alpha$}
\newcommand{\Hb}{$\rm H\beta$}
\newcommand{\hi}{{H{\sc i}}}
\newcommand{\hii}{{H{\sc ii}}}
\newcommand{\nii}{\ion{N}{2}}
\newcommand{\rband}{{\em r}-band}
\newcommand{\iband}{{\em I}-band}
\newcommand{\zband}{{\em z}-band}
\newcommand{\rd}{$r_{\rm d}$}
\newcommand{\whi}{$W_{50}$}
\newcommand{\ds}{$\Delta s$}
\newcommand{\x}{$\times$}
\newcommand{\about}{$\sim$}
\newcommand{\Msun}{M$_\odot$}
\newcommand{\Lsun}{L$_\odot$}
\newcommand{\Mhi}{$M_{\rm HI}$}
\newcommand{\Mst}{$M_\star$}
\newcommand{\must}{$\mu_\star$}
\newcommand{\nuvr}{NUV$-r$}
\newcommand{\Rinz}{$R_{50,z}$}
\newcommand{\Ropt}{$R_{\rm opt}$}
\newcommand{\sov}{$S_{0.5}$}
\newcommand{\vrot}{$V_{\rm rot}$}
\newcommand{\vs}{$V_{\rm rot}/\sigma$}
\newcommand{\cindx}{$R_{90}/R_{50}$}
\newcommand{\rhalf}{$R_{50}$}

\maketitle

\label{firstpage}

\begin{abstract}
We present dynamical scaling relations for a homogeneous and
representative sample of \about 500 massive galaxies, selected only by
stellar mass ($> 10^{10}$ \Msun) and redshift ($0.025<z<0.05$)
as part of the ongoing GALEX Arecibo SDSS Survey. We compare baryonic
Tully-Fisher (BTF) and Faber-Jackson (BFJ) relations for this sample, and
investigate how galaxies scatter around the best fits obtained for
pruned subsets of disk-dominated and bulge-dominated systems.
The BFJ relation is significantly less scattered than the BTF when the
relations are applied to their maximum samples (for the BTF, only
galaxies with \hi\ detections), and is not affected by the
inclination problems that plague the BTF. Disk-dominated, gas-rich
galaxies systematically deviate from the BFJ relation defined by the
spheroids. We demonstrate that by applying a simple correction to the
stellar velocity dispersions that depends only on the concentration
index of the galaxy, we are able to bring disks and spheroids onto the
same dynamical relation --- in other words, we obtain a generalized
BFJ relation that holds for all the galaxies in our sample, regardless
of morphology, inclination or gas content, and has a scatter smaller
than 0.1 dex. We compare the
velocity-size relation for the three dynamical indicators used in this
work, \ie, rotational velocity, observed and concentration-corrected
stellar dispersion. We find that disks and spheroids
are offset in the stellar dispersion-size relation, and that the
offset is removed when corrected dispersions are used instead. The
generalized BFJ relation represents a fundamental correlation between
the global dark matter and baryonic content of galaxies, which is
obeyed by all (massive) systems regardless of morphology. 
\end{abstract}


\section{Introduction}\label{s_intro}

The observed global properties of galaxies obey a diverse set of
scaling relations, which are fundamental tools to constrain models of
galaxy formation and evolution. Particularly interesting are the
correlations between dynamics and luminosity/stellar mass or size, such as
the Tully-Fisher \citep[TF;][]{tf77} relation for spirals, and the
Faber-Jackson \citep[FJ;][]{fj76}, $D_{\rm n}$-$\sigma$ \citep{dressler87a},
or fundamental plane \citep[FP;][]{dd87,dressler87b} relations for spheroids, 
because they link the luminous to the total mass of galaxies, thus
providing insights into the interplay between their baryonic and dark
matter components. 

Historically, the importance of the tight TF and FP relations as secondary
distance indicators has driven the community to assemble galaxy
samples that meet strict selection criteria, in order to minimize
systematic errors and scatter. While FP studies targeted
elliptical and S0 galaxies, the samples used for TF applications have
typically been restricted to late-type spirals with inclinations to
the line-of-sight larger than 30-40\deg, preferably observed in red
or infrared photometric bands to minimize extinction effects 
\citep[\eg,][]{courteau97,giovanelli97b,masters06}.
These data sets are not ideal for characterizing the statistical
properties of galaxies in general, because the excessive pruning means
that they are not fair samples of the local Universe. This issue is
particularly problematic for the comparison with theoretical studies
and numerical simulations of galaxy formation and evolution,
which should be based on representative samples.
For instance, TF studies of different classes of objects, such as
S0s and early-type spirals 
\citep[\eg,][and references therein]{eyal99,bedregal06,williams10}, 
polar ring galaxies \citep{iodice03}, barred spirals \citep{courteau03},
and gas-rich dwarfs \citep{mcgaugh00,begum08}, have sometimes
found disagreement with the TF relation of late-type spirals.
Most notably, gas-rich dwarfs lie systematically below the TF relation
defined by bright galaxies.

During the last decades, the interest in TF and FJ-like scaling
relations has shifted from cosmic flow applications to constraining
galaxy formation models, and samples with broader morphological
properties have been constructed specifically for this purpose
\citep[\eg,][]{kannappan02,pizagno07,dutton07,avila-reese08}. However, current large and
homogeneous data sets include either spirals (\eg, in addition to the works
mentioned above, \citealt{courteau07a} and \citealt{saintonge11}, hereafter SS11)
or early-type galaxies \citep[\eg,][]{bernardi03,graves09} only.

There have been attempts to move beyond the spiral/elliptical
dichotomy, and uncover relations between stellar content and dynamics that hold
for all galaxies, independent of morphology. \citet{zaritsky08} found that
all galaxies, from disks to spheroids and from dwarf spheroidals to
giant ellipticals, lie on a two-dimensional surface defined by surface
brightness, half-light radius, internal velocity and mass-to-light
ratio. As a measure of internal velocity $V$, they adopt 
either the rotational velocity \vrot\ for disks or the velocity
dispersion $\sigma$ for spheroids, thus the two types of systems are
still treated separately (especially because the sample is a large but
heterogeneous collection of published data sets, for which
either \vrot\ or $\sigma$ is available).

It is also important to point out that the search for scaling
relations that are valid for all types of galaxies should make use 
of {\it baryonic} masses (\ie, the sum of stellar and gas masses)
instead of luminosities or stellar masses, because they could be more
fundamental quantities. This is demonstrated by the fact that
baryonic scaling relations hold for subsets of galaxies that do not
follow the corresponding stellar relations. For example, the offset of
the gas-rich dwarf galaxies from the stellar TF relation disappears
when their gas mass is taken into account
\citep{mcgaugh00,begum08}. The baryonic TF relation
\citep[BTF;][]{mcgaugh00} is linear over 5 orders of magnitude in
(stellar + gas) mass, suggesting that the TF is fundamentally
a relation between baryonic (rather than luminous) and total mass of
the galaxy. Intriguingly, although supported by limited statistics,
there is some evidence that giant and dwarf ellipticals might lie
on the same BTF as the spirals \citep{derijcke07}.
Unfortunately, because they require estimates of both
stellar and gas masses, BTF samples 
\citep[\eg,][]{mcgaugh05,geha06,begum08,gurovich04,gurovich10}
are significantly smaller than TF ones.

To summarize, it is still unclear if disk-dominated galaxies and
spheroids obey the same dynamical scaling relations, mainly due to the
lack of well-defined, representative samples of galaxies for which both
rotation and stellar dispersion are measured. Certainly, ellipticals
might have no gas and no detectable rotation, and pure disks might have
negligible stellar dispersions, but a significant fraction of local
galaxies (especially massive ones) have a disk {\it and} a bulge
\citep[\eg][]{driver07}, and the dynamical scaling relations should account for the smooth
transition across galaxies with different bulge-to-disk ratios.
As \citet{covington10} point out, a connection between the scaling
relations of early-type and late-type galaxies is expected on the
grounds that early-type systems are generally assumed to form through
mergers of late-type ones. This is especially true at the high 
stellar mass end, where the blue sequence of star-forming disks
merges onto the red sequence of passively-evolving, bulge-dominated galaxies 
\citep[\eg][]{baldry06}, and the systems typically host a bulge and a disk.

In this paper, we investigate dynamical scaling relations for a
representative sample of \about 500 massive galaxies that are selected 
{\it only} by stellar mass ($M_\star > 10^{10}$ \Msun) and redshift
($0.025<z<0.05$), as part of the ongoing GALEX Arecibo SDSS Survey
\citep[GASS;][hereafter Paper I]{gass1}. For these galaxies, we have
homogeneous measurements of structural parameters and velocity
dispersions from the Sloan Digital Sky Survey \citep[SDSS;][]{sdss},
\nuvr\ colours from GALEX \citep{galex} and SDSS imaging, and
\hi\ masses and rotational velocities for the
subset of objects detected at 21 cm with the Arecibo radio telescope.
This unique sample, which
includes massive galaxies of all morphological types, allows us to
investigate how objects that are typically not included in TF or FJ/FP
data sets scatter around those relations. In the spirit of works like
those of \citet{zaritsky08} and \citet{covington10}, we wish to
establish if there is a {\it fundamental} correlation between
baryonic mass and dynamics that is obeyed by the complete galaxy
population, regardless of morphology.
We show that, at least for the massive galaxies in our sample,
such a relation does exist, and has a scatter smaller than 0.1 dex,
comparable to that of the TF and FJ relations applied to their
respective pruned subsets.

This paper is organized as follows. We summarize sample selection and
measurements of relevant quantities in \S~\ref{s_sample}. We present
the baryonic mass-velocity relations, starting with the TF and FJ, in 
\S~\ref{s_bvrel}, and the velocity-size relations in \S~\ref{s_vsize}.
We discuss our findings and conclude in \S~\ref{s_disc}.

All the distance-dependent quantities in this work are computed
assuming $\Omega=0.3$, $\Lambda=0.7$ and $H_0 = 70$ \kmsm.

\section{Sample Selection and Galaxy Parameters}\label{s_sample}

The sample used in this work is drawn from GASS, an on-going survey
which is gathering high-quality \hi -line spectra for \about
1000 massive galaxies, selected only by stellar mass (greater than
$10^{10}$ \Msun) and redshift ($0.025 < z < 0.05$). The GASS targets
are located within the intersection of the footprints of the SDSS
primary spectroscopic survey, the projected GALEX Medium
Imaging Survey and the on-going \hi\ blind Arecibo Legacy
Fast ALFA \citep[ALFALFA;][]{alfalfa} survey. The galaxies are 
observed with the Arecibo radio telescope until detected or until a gas
fraction limit of $1.5-5$\% is reached. For more details we refer the
reader to Paper I, where the first GASS data release (DR1) is presented.

Here, in addition to the DR1 data (176 galaxies), we use new Arecibo observations of
240 galaxies that will be incorporated in the second GASS data release
(Catinella et al., in preparation). 
As discussed in Paper I, GASS does not re-observe objects with good
\hi\ detections already available from the ALFALFA survey or the Cornell
\hi\ archive \citep[][hereafter S05]{s05}. To correct the GASS sample
for its lack of \hi-rich objects, we add galaxies from ALFALFA and S05
in the correct proportions, following the procedure detailed in
section 7.2 of Paper I. The sample thus obtained, which is representative
in terms of \hi\ properties, includes 480 galaxies (296 detections and 184
non-detections). As explained below, we discard 44 galaxies for which
the stellar velocity dispersion is not reliable. The final sample
includes 436 galaxies (259 detections and 177 non-detections).

As mentioned in Paper I, the optical parameters are obtained from
queries to the SDSS DR7 \citep{sdss7} data base server, unless otherwise
noted. Stellar masses are derived from SDSS photometry using the
spectral energy distribution (SED) fitting technique described in
\citet{salim07}, assuming a \citet{chabrier03} initial mass function. 
A variety of model SEDs from the \citet{bruzual03} library are fitted
to each galaxy, building a probability distribution for its stellar
mass. The mean and the width of this distribution are used as
measurements of the stellar mass and its formal error,
respectively. Over the interval probed by GASS, stellar masses
are believed to be accurate to better than 30 per cent.
As a measure of galaxy size we adopt $R_{25}$, \ie\ half the 
25 mag arcsec$^{-2}$ isophote diameter $D_{25}$ (measured by us on
the SDSS {\em g}-band images), in kpc.

The baryonic mass is the sum of stars and gas;
the latter is computed from the \hi\ mass, adopting the standard
1.4 correction factor to account for helium and metals, \ie\ 
$M_{\rm gas}= 1.4 M_{\rm HI}$.
This correction neglects the contribution of the molecular
hydrogen. In our stellar mass regime, the amount of $\rm H_2$ 
does not depend significantly on \Mst\ or concentration index and,
on average, $M_{\rm H_2}/M_{\rm HI}=0.30$ (but with a large scatter,
0.41 dex; \citealt{coldgass1}). Since the contribution of the \hi\ to
the baryonic mass is, for our sample, typically small (see \S~\ref{s_btf}),
it is safe to neglect the $\rm H_2$ (but we did check that including
the molecular gas does not change our results).
We set the \hi\ masses of the non-detections to zero, but we note that using
the upper limits (which correspond to 1.5-5\% of the stellar mass by
survey design) would make no difference to our plots. 
Other parameters used in this work
are discussed below.

\subsection{\hi\ line widths}\label{s_hiwidths}

\hi\ line widths from GASS, ALFALFA and the S05 archive are
measured with the same technique, at the 50\% of each peak level 
\citep[\eg,][\S 2.2]{widths}. However, the corrections applied to
the raw measurements, as originally published, are different. 
GASS and S05 widths are corrected for both instrumental broadening and
cosmological redshift, whereas ALFALFA widths are corrected for
instrumental broadening only, following equation~1 in \citet{kent08}.
No turbulent motion or inclination corrections are applied.
For convenience, we report here the corrections adopted:
\begin{displaymath}
        W_{\rm GASS, S05}^c = \frac{W_{50} -\Delta s}{1+z}
\end{displaymath}
\begin{displaymath}
        W_{\rm ALFALFA}^c = \sqrt{W_{50}^2 -(\Delta s)^2}
\end{displaymath}
\noindent
where \whi\ is the measured velocity width, $z$ is the galaxy
redshift, and \ds\ is the instrumental broadening correction, which differs for the
three sources. For GASS Paper I and ALFALFA, \ds\ is simply the final velocity
resolution of the spectrum after smoothing (i.e., between 5 and 21
\kms\ for GASS spectra, which are Hanning and boxcar smoothed, and
\about 10 \kms\ for ALFALFA spectra, which are Hanning smoothed only).
For S05, \ds\ equals $2 \Delta v_{\rm ch} \lambda$, where $\Delta v_{\rm ch}$ is the
channel separation in \kms\ and $\lambda$ is a complex function of
signal-to-noise ratio (SNR) and type of smoothing applied. For high
SNR, Hanning-smoothed spectra with $\Delta v_{\rm ch}=5$ \kms\ (\ie, the
ALFALFA case), $\lambda = 0.40$, thus \ds\ is of order of half the
velocity resolution after smoothing. This is smaller than the
correction adopted for GASS in Paper I, and larger than the ALFALFA
one, where the subtraction in quadrature makes the correction
negligible (i.e., $\leq 1$ \kms) for velocity widths larger than 50 \kms.

The issue of homogenizing velocity widths obtained from different sources
has been discussed in the past, and most recently by \citet{courtois09} for
the Extragalactic Distance Database \citep{tully09},
a large compilation of data for galaxy distance and peculiar motion
studies. We follow \citet{courtois09}, who adopt a simplified version
of the S05 solution, \ie:  $\Delta s = 2 \Delta v \lambda$, where
$\lambda = 0.25$ and $\Delta v$ is the final velocity
resolution of the spectrum after smoothing (this is half the
correction adopted in Paper I). This is in better agreement with our
own tests on high SNR GASS \hi\ profiles, where the smoothing was gradually
increased and the width remeasured.
Thus, we keep the S05 widths as published\footnote{
For the only galaxy in common between S05 and GASS (GASS 38751,
AGC 240702), the corrected velocity widths agree within the quoted errors.
},
we adopt the new \ds\ correction for the GASS data, and 
we uncorrect the ALFALFA widths, $W_{\rm ALFALFA}^c$, to obtain the
raw \whi\ measurements, then apply the same \ds\ and $(1+z)$ 
corrections adopted for GASS galaxies. 

Lastly, we deproject all velocity widths to edge-on view. The
inclination to the line-of-sight is computed from:
\begin{equation}
{\rm cos}~i = \sqrt \frac{(b/a)^2 - q_0^2}{1-q_0^2},
\end{equation}
\noindent
where $b/a$ is the minor-to-major axis ratio from the 
\rband\ exponential fit ({\it expAB\_r} in the SDSS database),
and $q_0$ is the intrinsic axial ratio of a galaxy seen edge-on. 
We adopt  $q_0=0.20$ (\citealt{holmberg58}; see also, \eg\ \citealt{tully09}), 
and set the inclination to 90\deg\ for galaxies with $b/a < 0.2$.
The value of $q_0=0.20$ applies to galaxies of morphological type
earlier than Sbc, whereas $q_0=0.13$ should be used for Sbc and later
spirals (\eg\ \citealt{giovanelli94,giovanelli97a}; SS11).
However the difference is small, and our choice of $q_0$ seems
appropriate for a sample of massive galaxies that was not selected for TF studies.

Thus, rotational velocities are computed as:
\begin{equation}
        V_{\rm rot} = \frac{W_{50} - 0.5 \Delta v}{2~(1+z)~{\rm sin}~i},
\label{eq_vrot}
\end{equation}
where $\Delta v$ is the final velocity resolution of the spectrum
after smoothing.

As noted in Paper I, the \hi\ masses of galaxies in the ALFALFA and
S05 archives have been recomputed from the tabulated fluxes, in order
to be made consistent with GASS ones.

We identified 31 galaxies for which \hi\ confusion within
the \about 4\arcmin\ Arecibo beam is certain, and therefore the
measured \hi\ parameters should not be trusted. Specifically, we
closely inspected the SDSS images of all the galaxies
in our sample, and flagged as ``confused'' those with at least one
late-type, similar size companion (based on SDSS spectroscopy,
i.e. with redshift available and within 0.002 of that of the target
galaxy) within the beam. These objects will be highlighted in our
analysis when needed.

\subsection{Velocity dispersions}\label{s_vdisp}

Galaxy velocity dispersions are measured by fitting stellar templates
convolved with Gaussian functions to SDSS spectra, which are obtained 
through 3\arcsec -diameter fiber apertures. Because the fiber covers
only a fraction of the galaxy light at the GASS redshifts, these 
quantities (catalogued as {\it velDisp} in the SDSS spectroscopic data
base, and here referred to as $\sigma_{\rm fib}$) need to be corrected for
aperture effects.
As commonly done \citep[see, \eg][and references therein]{bernardi03,graves09}, 
fiber velocity dispersions are corrected to 1/8 effective radius as follows:
\begin{equation}
   \sigma = \sigma_{\rm fib} \left( \frac{r_{\rm fib}}{r_0/8} \right) ^{0.04}
\label{eq_sigma}
\end{equation}
\noindent
where $r_{\rm fib}=1.5\arcsec$ and $r_0$ is the circular galaxy radius in
arcseconds, computed as $r_0 = R_{\rm deV} \sqrt{(b/a)_{\rm deV}}$ ($R_{\rm deV}$
and $(b/a)_{\rm deV}$ are the effective radius and axis ratio from the
\rband\ de Vaucouleurs fit, respectively). This correction is small, 
\about 3\% on average, for the galaxies in our sample.

\begin{figure}
\includegraphics[width=7cm]{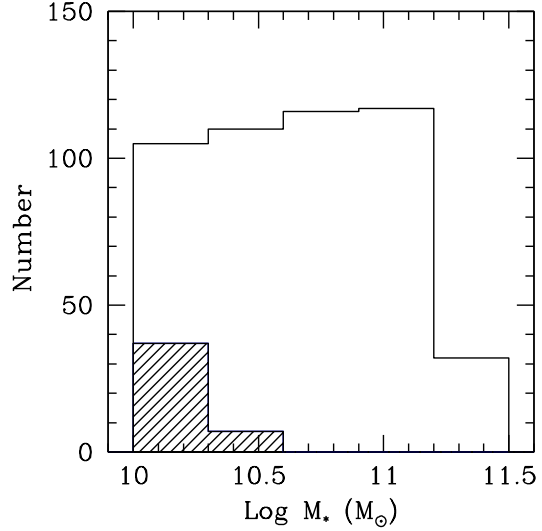}
\caption{Stellar mass distribution for the GASS sample (solid). The
shaded histogram shows the 44 objects with stellar velocity dispersion
smaller than 70 \kms\ that are not included in our analysis.}
\label{mstar}
\end{figure}

Only values of $\sigma_{\rm fib}$ above the \about 70 \kms\ instrumental resolution of the
SDSS spectra are considered reliable. We thus discarded 44 galaxies that
do not meet this requirement from our sample; the effect of this
restriction on the stellar mass distribution of the sample is shown in
Figure~\ref{mstar}.

\section{Baryonic Mass-Velocity Relations}\label{s_bvrel}

The main goal of this work is to establish if there is a relation
between baryonic mass and a measure of the dynamical mass (estimated based on
rotational velocity, stellar dispersion or a combination of the two)
that is obeyed by massive galaxies {\it regardless of their morphology}.

We begin by comparing the baryonic TF (BTF) and baryonic FJ (BFJ) relations for our sample, 
in order to establish which
quantity, the \hi\ rotational velocity or the stellar dispersion, most
reliably traces the baryonic mass of massive galaxies. Thanks to its unique selection
by stellar mass and redshift only, the GASS sample includes massive galaxies 
of all morphological types, and it is thus ideal for this comparison.
Indeed, this is not a sample designed for either TF or FJ 
studies, but we can prune it to separately study inclined, disk-dominated and 
spheroidal, bulge-dominated systems. Most importantly, we can investigate
outliers and residuals of the BTF and BFJ relations, which is essential in order 
to determine how to bring disks and spheroids onto the same relation. We conclude by
illustrating two different ways of obtaining a baryonic relation that
holds for all the massive galaxies in our sample, and whose
dispersion is comparable to that of the BTF and BFJ relations.

\begin{figure*}
\includegraphics[width=17cm]{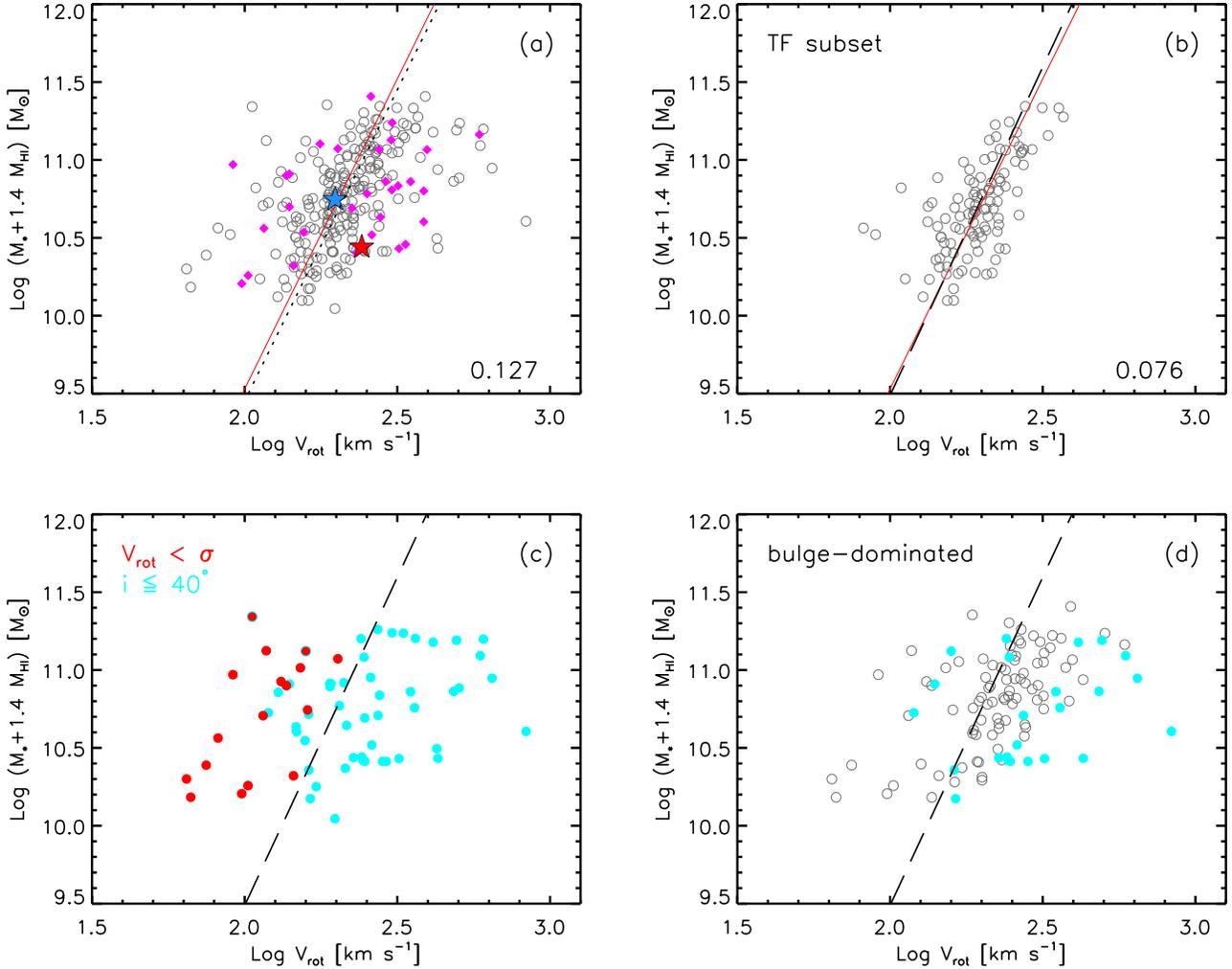}
\caption{Relation between baryonic mass and rotational
  velocity for all the galaxies with \hi\ detections (a). Magenta
  diamonds indicate objects for which \hi\ confusion is certain; red
  and blue stars correspond to GASS 3505 and 35981, respectively (see text). 
  The other panels show the same relation for selected subsets: (b)
  inclined, disk-dominated systems, (c) galaxies with inclinations
  smaller than 40\deg\ (cyan) and/or with $V_{\rm rot} < \sigma$ (red), and
  (d) bulge-dominated objects (filled cyan circles are 
  bulge-dominated galaxies with inclination smaller than 40\deg).
  In the top panels, a dotted (a) or 
  dashed (b) line indicates the inverse fit to the data points (the
  scatter is noted in the bottom right corner), and the red solid line
  shows the BTF relation from \citet{mcgaugh00}. The dashed line in
  (b) is reproduced in panels (c) and (d) for reference.}
\label{btf}
\end{figure*}

\subsection{Baryonic Tully-Fisher Relation}\label{s_btf}

Here we must restrict our sample to galaxies with \hi\ detections, for which 
the rotational velocity can be measured. The BTF relation for this sample is shown in
Figure~\ref{btf}a. We note that, for our interval of
stellar masses, the contribution of the gas to the baryonic mass is
generally small --- if we plotted stellar instead of baryonic masses,
most of the points would move downward by an amount comparable to the
symbol size (indeed, there are only 13 galaxies in our sample with gas
fractions \Mhi/\Mst $>50\%$).
We plot the BTF as usually done (\ie, baryonic mass versus rotational velocity), 
but we fit the inverse relation, since the scatter is clearly associated 
to the \vrot -coordinate (as can be seen by comparing Figures~\ref{btf}a and
\ref{btf}b, explained below).  The inverse fit to the data points, \ie\
Log \vrot = a Log (\Mst + 1.4 \Mhi) + b, is shown as a dotted line\footnote{
Tables~\ref{t_fits} and \ref{t2_fits} list the expression of the fit, the scatter, and
the number of galaxies contributing to the sample for
all the main relations discussed in this work.
}.
Following SS11, we compute the scatter in the $x$
variable around the best fit by applying Tukey's bi-weight, which
yields a robust estimate of the dispersion in presence of outliers;
the scatter (in dex of velocity) is noted at the bottom of the
panel. Also shown is the relation obtained by \citet{mcgaugh00}
over nearly 5 orders of magnitude of baryonic mass (solid line), which
is in excellent agreement with our data.

Not surprisingly, the scatter around the best fit is large for
our data set (0.127 dex), and discarding galaxies with confused
\hi\ spectra (magenta symbols; see \S~\ref{s_hiwidths}) improves it
only by a small amount (0.112 dex). However, we recover a significantly tighter
relation when we prune the sample as usually done for TF studies. This
is illustrated in Figure~\ref{btf}b, where a {\it TF subset} was
obtained by selecting non-confused, disk-dominated, inclined galaxies
(\ie, objects with concentration index \cindx $\leq 2.8$ and
inclination $i>40$\deg; the scatter of the BTF for this subset is 0.076 dex).
We inspected the three low-velocity outliers that stand out on the
left of the plot. Two of them are galaxies for which the SDSS inclination
is clearly overestimated (we measured 17\deg\ and 26\deg\ instead of
46\deg\ and 49\deg, respectively), and the third one has uncertain
\hi\ velocity width. Removing these outliers, the scatter of the
relation becomes 0.072 dex. How does this compare with the scatter of
other TF samples? There is a vast literature on the TF relation, showing that
its scatter depends on the sample analyzed, velocity indicator,
photometric band, and type of fit performed (forward, inverse,
bisector or orthogonal). \citet{mcgaugh05} studied the BTF relation
using a sample of 60 galaxies with extended \hi\ rotation curves, and
obtained a scatter of 0.191 dex in solar masses from a direct fit
(this is the average scatter for the relations marked as $\cal P^{\rm d}$ 
in his Table 2, where the stellar masses are computed using stellar
population synthesis models as we do). For comparison, a direct fit to 
the data in Figure~\ref{btf}b (excluding the three low-velocity outliers mentioned
above) yields a scatter of 0.221 dex in solar masses.
A study that is directly comparable to ours
is that of \citet{avila-reese08}, who measured a scatter of 0.06 dex
in velocity for the BTF of a sample of 76 non-interacting disk galaxies, with
inclinations $35^{\circ} \leq i \leq 80^{\circ}$ and flat rotation curves
\citep[see also, \eg,][]{pizagno07}.
Given the crude selection of our TF subset (by concentration index),
our sample is likely to include a broader range of disk galaxy types,
thus the slightly larger scatter compared to these studies is not unexpected.

Figure~\ref{btf}c demonstrates that the high-velocity outliers of the
BTF relation are all galaxies with small inclination to the line-of-sight.
For these systems, the corrected rotational velocities become so
unreliable that the correlation with baryonic mass effectively
disappears. Once galaxies with inclination smaller than 40\deg\ are
removed from the sample, even bulge-dominated objects lie reasonably
close to the relation obtained for disk-dominated ones, although with
larger scatter (empty circles in Fig.~\ref{btf}d).
This agrees qualitatively with the findings of \citet{ho07}, 
based on a compilation of 792 galaxies with inclination larger than
30\deg\ and heterogeneous measurements from Hyperleda 
\citep{paturel03a,paturel03b}. Indeed, Ho showed that a $K_S$-band TF
relation exists for all Hubble types, including elliptical and S0
galaxies (although these represent a very small fraction of his sample). 

Figure~\ref{btf}c also shows a population of low-velocity outliers
that cannot be explained by small inclinations: these are galaxies
with rotational velocities that are smaller than the stellar velocity
dispersions (red symbols).  We note that \citet{ho07} 
also identifies a population of galaxies with unusually small \vs\
ratios, which are outliers in his TF relation. Except for their small
rotational velocities compared to their central stellar velocity
dispersions, his outliers are otherwise normal luminous galaxies. 
These systems might be very interesting --- Ho argues that a
significant fraction of their \hi\ gas must be
dynamically unrelaxed, having been acquired through a minor merger
episode or perhaps cold accretion. For our sample, the fraction of low
\vs\ outliers is smaller (\about 5\%, against 17\% for Ho's
sample). As shown in the Appendix, where these galaxies are
described in more detail, most of them have asymmetric
\hi\ profiles, suggesting that the \hi\ distribution and/or
kinematics might be disturbed (although we cannot prove that the
gas was externally accreted).

Lastly, here and in other figures of this paper we highlight the
positions of two interesting galaxies that were discussed in our
previous works, GASS 3505 and 35981 (marked as a red and a blue star
in Fig.~\ref{btf}a). These are both unusually \hi-rich systems, which are
outliers in the gas fraction plane discussed in Paper I (relating the
\hi\ mass fraction to stellar mass surface density \must\ and
\nuvr\ color). However, GASS 3505 is an early-type system, whereas
GASS 35981 (UGC 8802) is a disk galaxy with a sharp metallicity drop 
in its outer disk, which suggests an external origin
for the \hi\ gas (\citealt{moran10}; see also \citealt{wang11}). 
As can be seen, from a dynamical point of view GASS 35981 is not
unusual, whereas GASS 3505 is still an outlier.

\begin{figure*}
\includegraphics[width=17cm]{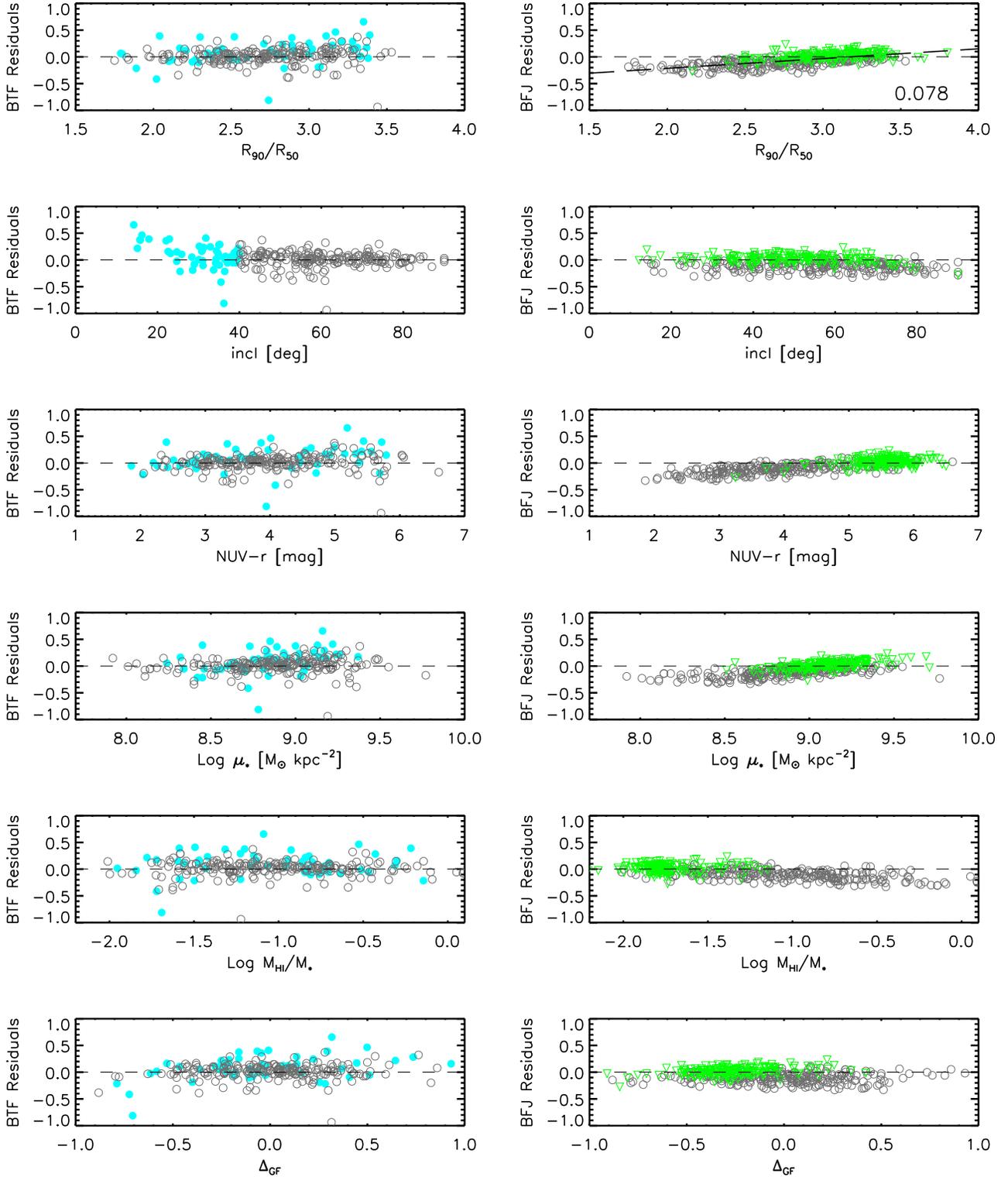}
\caption{Residuals of baryonic TF (left column) and FJ (right column) 
  relations plotted as functions of concentration index
  (row 1), galaxy inclination (row 2), \nuvr\ color corrected for
  Galactic extinction only (row 3), stellar surface density (row 4),
  gas fraction (row 5), and distance from the gas fraction plane (see
  text; row 6). Filled cyan circles and green upside-down triangles
  indicate galaxies with inclinations smaller than 40\deg\ and \hi\ non-detections, respectively.
  For the top right panel, we show a linear fit to the data
  points (long-dashed line) and note its dispersion on the bottom right
  corner (see text).}
\label{residuals}
\end{figure*}

We now investigate how the residuals of the BTF relation depend on a few
representative galaxy parameters. The left column of
Figure~\ref{residuals} shows the BTF residuals for
the sample in Figure~\ref{btf}a with respect to the best inverse
fit obtained for the {\it TF subset} in Figure~\ref{btf}b. In other
words, residuals are computed as $Log~x - Log~x_{\rm fit}$, where $x$
is the measured rotational velocity and $x_{\rm fit}$ is the value
expected from our best BTF relation for a galaxy with the same
baryonic mass. Cyan filled circles indicate objects with inclination
smaller than 40\deg. From top to bottom, residuals are plotted as
functions of concentration index, inclination, \nuvr\ color, 
stellar mass surface density, gas fraction, and distance
from the gas fraction plane mentioned above and described in Paper I
(galaxies more \hi-rich than the average have positive distance),
respectively. BTF residuals do not exhibit strong dependence
on structural and star-forming galaxy properties, except for a mild
tendency toward increased scatter for more bulge-dominated, red
galaxies, which largely disappears when the systems with low
inclinations are removed from the sample. This is not surprising, as
many attempts to identify a third parameter to minimize the 
scatter of the TF relation produced negative results 
\citep[e.g.,][and references therein]{courteau99,pizagno07,meyer08}.

The results presented in this section show that the BTF is not very
promising for our purpose of determining a relation between baryonic
and dynamical mass that holds for all massive galaxies.
Firstly, its application is restricted to galaxies with \hi\
detections and inclinations larger than  40\deg. Secondly, the
bulge-dominated systems with \hi\ detections are not simply offset
from the relation defined by the disk-dominated galaxies, but are also
more scattered. As demonstrated in the following two sections, the BFJ
relation does not suffer from these limitations, and can be generalized to
include both disk-dominated and spheroidal systems.

\begin{figure*}
\includegraphics[width=17cm]{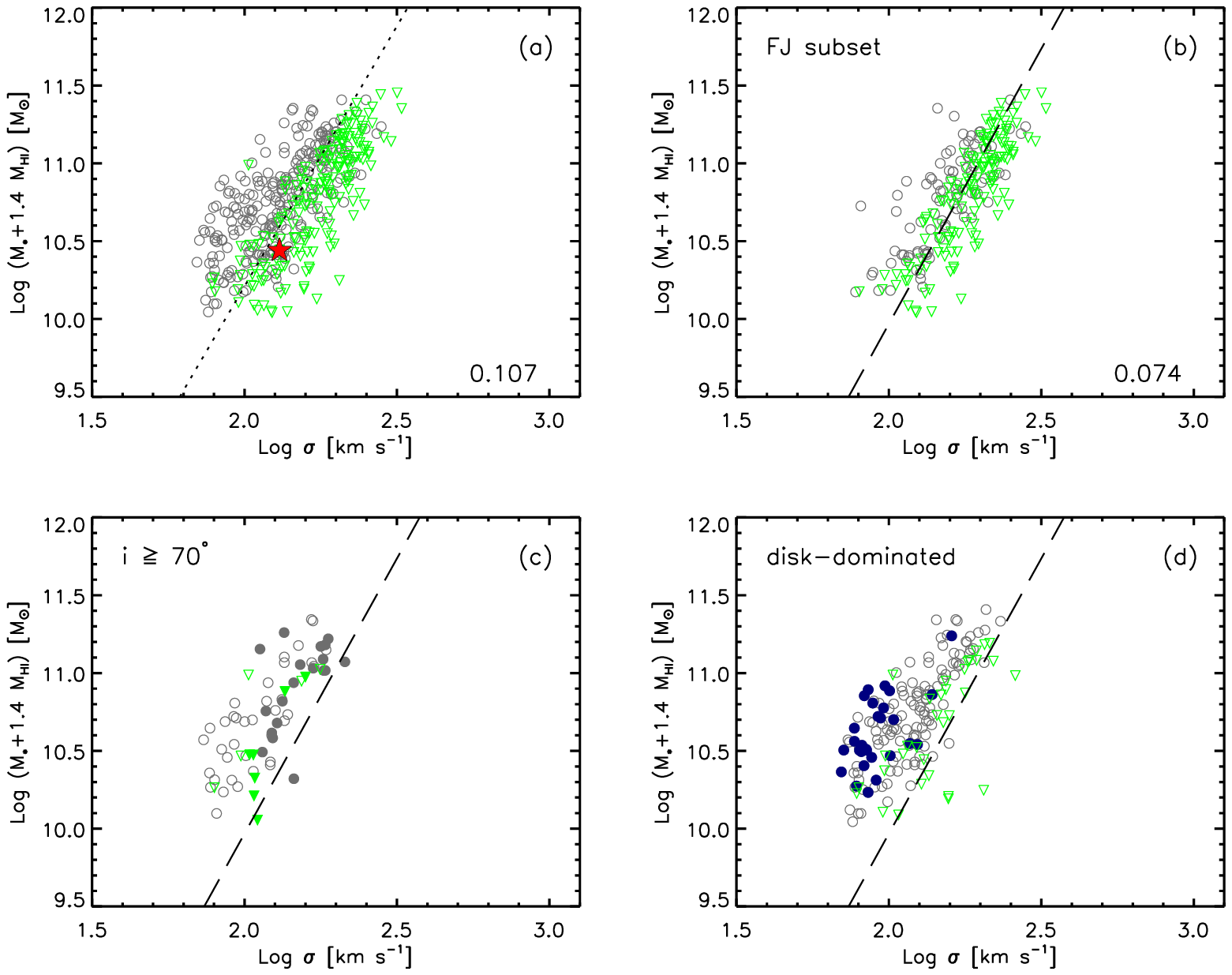}
\caption{Relation between baryonic mass and stellar
  velocity dispersion for our sample (a) and for selected subsets: 
  (b) bulge-dominated systems with inclinations smaller than 70\deg, 
  (c) galaxies with large inclinations, and (d) disk-dominated objects.
  Green upside-down triangles indicate galaxies that were not detected in \hi. 
  The red star in (a) is GASS 3505; filled symbols in (c) indicate galaxies that are both
  bulge-dominated and with high inclinations, and blue filled circles in (d) are
  disk-dominated galaxies with large gas fractions (\Mhi/\Mst $>$
  30\%). In the top panels, a dotted (a) or 
  dashed (b) line indicates the inverse fit to the data points (the
  scatter is noted in the bottom right corner). The dashed line in
  (b) is reproduced in panels (c) and (d) for reference.}
\label{bfj}
\end{figure*}

\subsection{Baryonic Faber-Jackson Relation}\label{s_bfj}

\begin{figure*}
\includegraphics[width=14cm]{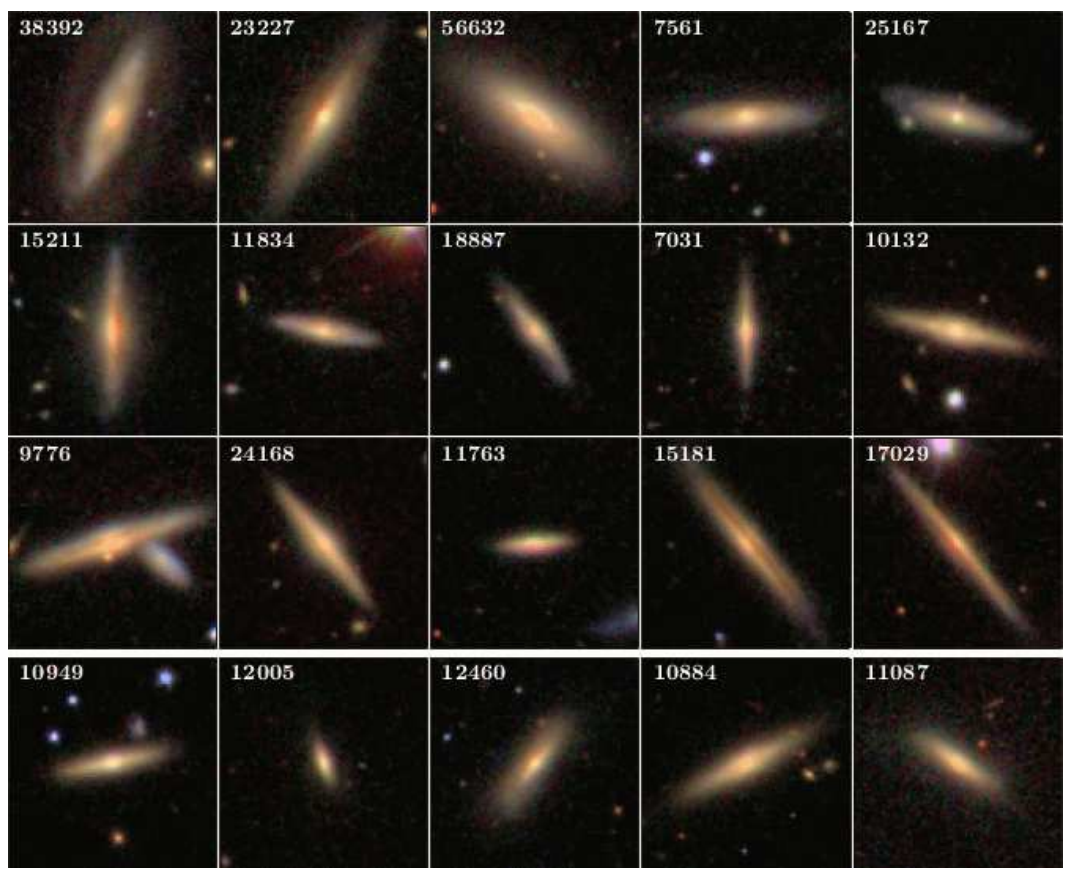}
\caption{SDSS postage stamps (1\arcmin\ size) for a subset of
  bulge-dominated galaxies with inclinations larger than 70\deg\ 
  shown in Figure~\ref{bfj}c (filled symbols). The galaxies are
  labeled with their GASS identifier. Top three rows: galaxies with
  \hi\ detections, ordered by increasing inclination. Bottom row:
  \hi\ non-detections, also ordered by increasing inclination.}
\label{bfj_edgeons}
\end{figure*}

We now carry out a similar analysis for the BFJ relation. Since all the galaxies
have a measurement of the stellar velocity dispersion from SDSS, we can 
use here the full sample, but it is instructive to keep \hi\ detections and
non-detections separated.
In Figure~\ref{bfj} the baryonic mass is plotted as a function of the
stellar velocity dispersion, and green upside-down triangles indicate
galaxies that were not detected with Arecibo.
The relation plotted for the full GASS sample (panel a) shows a clear 
segregation between \hi\ detections and non-detections, the former
being offset towards lower values of velocity dispersion. 
The gas-rich elliptical GASS 3505 (red star) is not unusual in terms
of its stellar dispersion (GASS 35981 is not plotted because it has 
$\sigma < 70$ \kms).
We recover a tighter relation when we restrict our sample to the 
{\it FJ subset} (panel b), which includes bulge-dominated galaxies (\ie,
objects with \cindx $> 2.8$) with inclinations smaller than
70\deg. We excluded from the subset highly flattened systems because
these are likely to host a significant disk component, despite their high
concentration index. The 26 bulge-dominated galaxies excluded by our
inclination cut are shown as filled symbols in Figure~\ref{bfj}c. These are mostly
\hi\ detections, which already suggests that they are more likely to
be disks than oblate spheroids. Their SDSS images
(Figure~\ref{bfj_edgeons}) confirm that they are inclined
disks, often with dust lanes running along their major axis.
As demonstrated by the bottom panels of Figure~\ref{bfj}, disk-dominated
galaxies are systematically offset from the best fit relation obtained
for the data points in panel (b), and they are mostly \hi\ detections. 
These objects have low stellar velocity dispersions for fixed baryonic
mass, thus they increase the scatter of the FJ relation when they are
plotted together with the bulge-dominated galaxies.

\begin{figure}
\includegraphics[width=8cm]{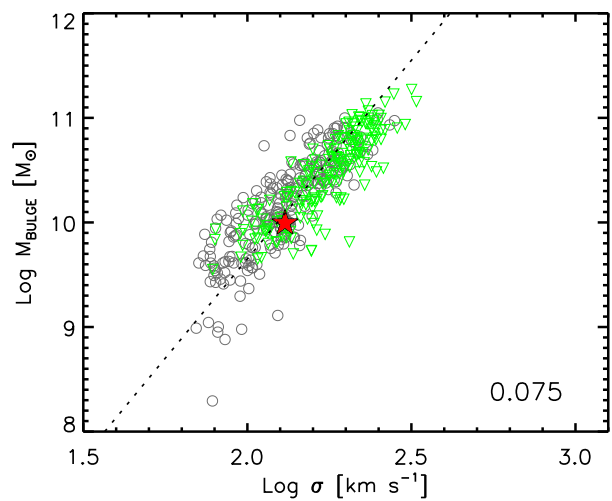}
\caption{Relation between bulge mass and velocity dispersion. Symbols
and dotted line as in Figure~\ref{bfj}a.}
\label{mbulge}
\end{figure}

Because FJ studies usually do not include the gas component, we
check our results by considering the {\it stellar} FJ relation.
In particular, we compared our FJ relation with the one obtained for a large sample of
SDSS galaxies by \citet{gallazzi06}, which is conveniently expressed
in terms of stellar masses. The slope and scatter of our
FJ relation (restricted to the {\em FJ subset}) are in remarkable
agreement with those published by \citet{gallazzi06} (both relations
have a scatter of 0.071 dex), whereas the zero
points differ by 0.17 dex in solar masses. The \citet{gallazzi06} sample
includes galaxies with concentration index $\geq2.8$, redshift
$0.005<z\leq 0.22$, and whose SDSS spectra have a median signal-to-noise
per pixel greater than 20. Removing our inclination cut does not bring
the zero points of the two relations into agreement (the offset is
still 0.14 dex). However, \citet{gallazzi05} note that their stellar
masses are systematically larger than those derived by
\citet{kauffmann03a} (and adopted here) by
\about 0.1 dex, thus our results are completely consistent.
Lastly, we notice that adding the gas to the stellar mass
slightly increases the scatter of the relation (from 0.071 dex for the
FJ to 0.074 dex for the BFJ). This is because the most
\hi-rich galaxies in our sample (see for instance the blue points 
in Fig.~\ref{bfj}d) become even further displaced from the
\hi\ non-detections, which constitute the bulk of the {\it FJ subset}.
A relation with similar scatter is obtained when, instead of
restricting the sample to bulge-dominated systems, we consider only the
bulge component itself. The mass of the bulge can be estimated 
from the correlation between SDSS concentration index
and bulge-to-total stellar mass ratio, B/T, presented by 
\citet[][see their Fig. 1]{weinmann09}, which is based on the 2D multicomponent decomposition
analysis of \citet{gadotti09}. Our bilinear fit to their data yields
B/T=(\cindx$-$1.920)/2.276. The relation between bulge mass and 
stellar velocity dispersion plotted in Figure~\ref{mbulge} has a
scatter of 0.075 dex in velocity, comparable to that of the FJ and
BFJ relations applied to the FJ subset \citep[see also][]{gadottikauffmann09}.
Incidentally, it is somewhat surprising that we can recover such a
tight relation given the uncertainties in our crude B/T estimate.

We now focus again on the
{\it baryonic} FJ relation presented in Figure~\ref{bfj} and study its residuals.
The BFJ residuals are plotted as functions of \cindx, inclination,
\nuvr\ color, stellar mass surface
density, gas fraction and distance from gas fraction plane
on the right column of Figure~\ref{residuals}. As for the BTF,
these are the residuals for the sample in Figure~\ref{bfj}a with
respect to the best fit relation in Figure~\ref{bfj}b, computed as
described in \S~\ref{s_btf} (where $x$ is now stellar velocity
dispersion). As in all the figures in this paper,
green upside-down triangles indicate \hi\ non-detections.
Interestingly, BFJ residuals show clear trends with concentration
index, \nuvr\ color, \must, and gas fraction, in the sense that more
disky, gas-rich and star-forming
galaxies have larger residuals. A similar trend is seen as a function of
D$_n$4000 index, which is an indicator of the age of the stellar
population sampled by the SDSS 3\arcsec -diameter fiber (larger
residuals are seen for smaller values of D$_n$4000, not shown). It is
well known that the FJ relation for
elliptical galaxies is a projection of a more general Fundamental Plane 
\citep{dd87,dressler87b}, which is usually parametrized in
terms of stellar velocity dispersion, effective radius $r_{\rm e}$, and
luminosity or surface brightness (central, or average within $r_{\rm e}$).
Thus, one might expect to see a dependence of the FJ residuals on a
third parameter related to $r_{\rm e}$. However, the trends 
in Figure~\ref{residuals} are {\em not} a consequence of the
Fundamental Plane. Firstly, our sample does not include
only elliptical galaxies and spheroids. As can be seen by inspecting
Figure~\ref{residuals}, the trends are driven by the disk-dominated,
star-forming galaxies. Secondly, quantities such 
as \cindx, \nuvr\ color or gas fraction are not directly related to an effective radius.

The comparison of the BTF and BFJ relations for our sample of massive galaxies
illustrates several important points:
(a) the BFJ is significantly less scattered than the BTF when the
relations are applied to their maximum sample. When the two relations
are applied to their respective ``good'', morphologically-pruned subsets, the
scatter in both is almost identical.
(b) the BFJ is insensitive to the inclination problems that plague the
BTF, which can be applied only to systems with inclination larger
than 40\deg. Furthermore, stellar dispersions are measured also for
galaxies without \hi\ detections. Naturally, one could measure
rotational velocities with other tracers and methods, \eg, with \Ha\
rotation curves. However, these have their own sets of problems (for instance,
\Ha\ emission is typically significantly less extended than \hi\
emission, thus it may not trace the full rotational velocity. Moreover,
it is unclear at which spatial position the rotational velocity should
be measured. See, \eg, \citealt{widths}), and the inclination issue remains.
(c) Most importantly, and contrary to the BTF case, the BFJ
residuals show systematic trends with other galaxy properties. As
shown in Figures~\ref{residuals} and \ref{bfj}, disk-dominated galaxies 
do not form a scatter plot in the BFJ plane, but they systematically
deviate from the main relation defined by the bulge-dominated systems.

One might wonder if, for disk-dominated, inclined galaxies, $\sigma$ is strongly
affected by rotation. In other words, the BFJ for highly inclined disks
(Figure~\ref{bfj}c and \ref{bfj}d) could just
be a TF relation in disguise. From the average rotation curves of disk
galaxies of \citet{templates}, we estimated that the typical rotational velocity 
reached at 1.5\arcsec\ by GASS objects is \about 100 \kms\ 
(but could be up to twice as large for the most luminous and/or most
distant galaxies in our sample)\footnote{
We converted SDSS model $r$ magnitudes into Cousins $I$ magnitudes
following \citet{highz}, and adopted $R_{90,i}$, the radius containing
90\% of the Petrosian flux in the SDSS $i$-band, as a proxy for 
the optical radius \Ropt, which is the radius encompassing 83\% of the total
\iband\ light of the galaxy. The average \iband\ absolute
magnitude for our sample is $-$22.35 mag, and the average coverage of
the SDSS fiber radius is 14\% \Ropt.
}.
However, this does not take inclination and 
bulge-to-disk ratio within the aperture into account. Most importantly, 
the fiber measurements are intensity-weighted, thus most of the disk
stars contributing to the dispersion are those closer to the galaxy
center, where \vrot\ is negligible. Thus we conclude that $\sigma$
measurements are not significantly affected by contamination from disk
stars in circular orbits.

In summary, the SDSS stellar velocity dispersions provide an estimate of dynamical mass that 
--- at least in the stellar mass and redshift regime probed by GASS ---
is not only more generally applicable (\ie, not limited to inclined
galaxies with \hi\ detections or with extended \Ha\ rotation curves),
but also less affected by measurement problems (which are largely
responsible for the BTF outliers) compared to \vrot.
Moreover, the fact that the disk-dominated galaxies
follow a BFJ relation that is simply offset from that of the bulge-dominated
systems suggests a simple way of bringing the two onto the same
relation. This is accomplished in the next section.

\begin{figure}
\includegraphics[width=8cm]{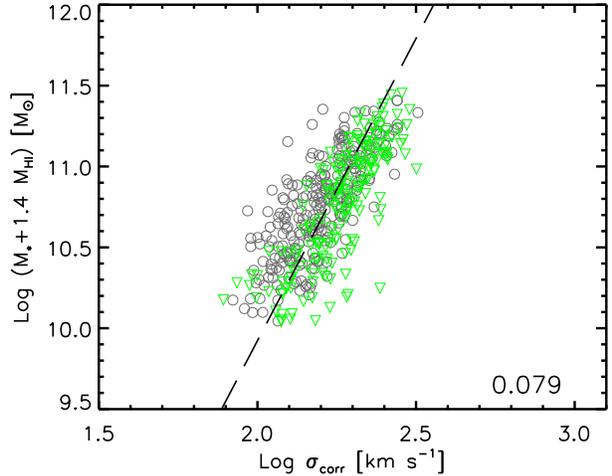}
\caption{Empirical baryonic FJ relation for massive galaxies, obtained by
  correcting the velocity dispersions for the systematic dependence on
  \cindx\ shown in Figure~\ref{residuals} (top right panel). As in
  other figures, \hi\ non-detections are indicated as green triangles.}
\label{bfj_corr}
\end{figure}

\subsection{A Generalized Baryonic Faber-Jackson Relation for All Massive Galaxies}\label{s_mbfj}

Spheroids and disk-dominated galaxies can be brought onto the same
BFJ-like relation by correcting the velocity dispersions for the
trends observed in  Figure~\ref{residuals}. The tightest dependencies
of the BFJ residuals are seen as a function of \nuvr\ color and
concentration index. Although the correlation with \nuvr\ is slightly
tighter (the dispersion of the linear fit is 0.073 dex), we decided to 
correct for the dependence on \cindx, which is a galaxy structural
parameter. In fact, it is reasonable to expect larger
corrections for more disk-dominated objects, where rotation is
likely to be the main contributor to dynamical support.
On the other hand, \nuvr\ color is linked to the star formation
properties of the galaxy, thus the rationale for correcting velocity
dispersions based on this quantity is much less evident. Moreover, a
correction based on \nuvr\ has the additional disadvantage of
requiring UV photometry.

\begin{figure}
\includegraphics[width=8cm]{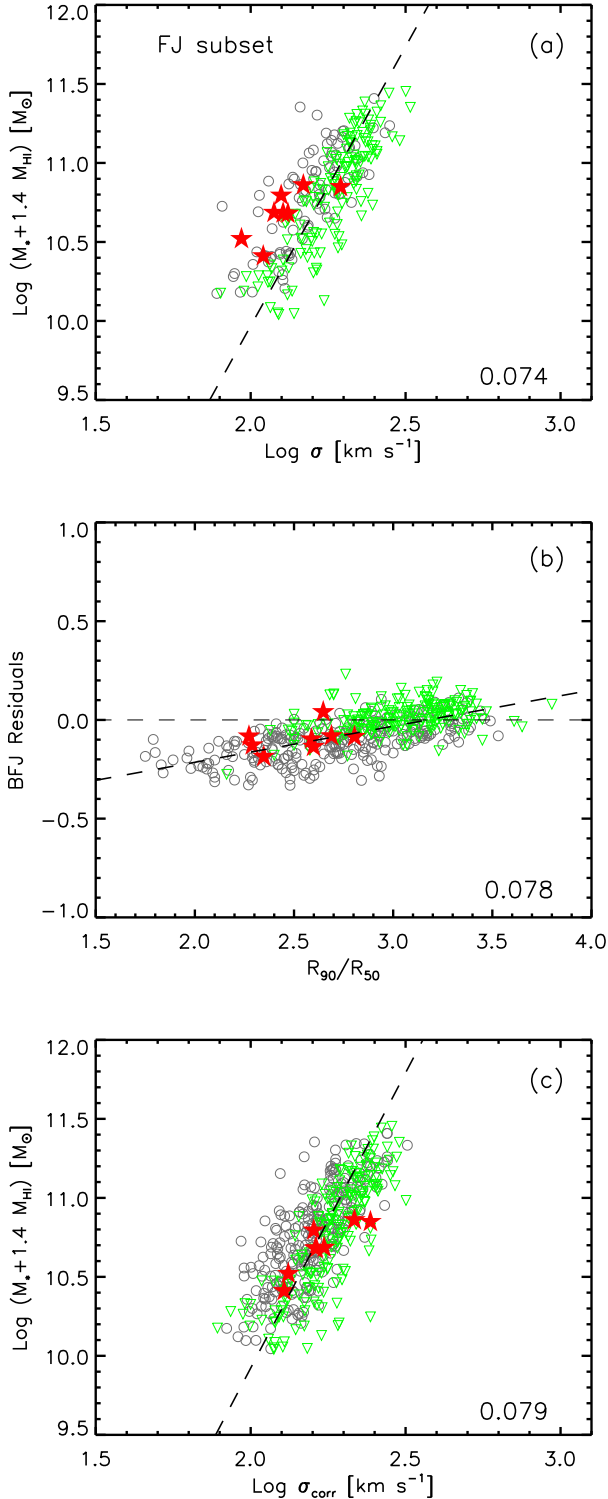}
\caption{Comparison between data (circles and triangles, corresponding
to \hi\ detections and non-detections, respectively) and simulations
(red stars, see text): BFJ relation (a), BFJ residuals as function of
\cindx\ (b), and generalized BFJ relation (c).}
\label{bfj_10kpc}
\end{figure}

The best linear fit to the data points in the top right panel of 
Figure~\ref{residuals} is:
\begin{equation}
    \Delta \sigma = -0.580 + 0.183 ~R_{90}/R_{50}
\label{eq_modsigma}
\end{equation}
\noindent
where the residuals $\Delta \sigma = Log ~\sigma - Log ~\sigma_{\rm BFJ}$
are computed with respect to the best fit BFJ relation in
Figure~\ref{bfj}b. We thus corrected the velocity dispersions 
$Log~\sigma$ of all the galaxies in our sample by subtracting the
offset $\Delta \sigma$. The BFJ relation obtained using the
corrected velocity dispersions is plotted in Figure~\ref{bfj_corr}.
The inverse fit is almost indistinguishable from that of
Figure~\ref{bfj}b, and has a scatter of 0.079 dex (the scatter in
solar masses, obtained from a direct fit, is 0.218 dex). This relation holds
for all the galaxies in our sample, regardless of morphology,
inclination or gas content, and is the main result of this work.

We compared our results with the simulations performed by 
\citet{scannapieco09,scannapieco11}, which follow the
formation of eight Milky Way-mass halos in a $\Lambda$-cold dark
matter cosmology, including baryonic physics (star formation, metal
cooling, chemical enrichment, multiphase gas, thermal feedback from
supernovae). For these simulated galaxies we can measure stellar and
cold gas masses, and stellar velocity dispersions at a given radius,
and calculate concentration indices (from the ratio of the radii
enclosing 90\% and 50\% of the total luminosity; we used
\rband\ luminosities computed from the dust-free \citealt{bruzual03}
population synthesis models). We estimate total masses and
luminosities within 10 kpc (the mean value of $R_{90}$
for the galaxies in the GASS sample), and obtain luminosity-weighted
stellar velocity dispersions within 1.5 kpc (the physical size
subtended by a 1.5\arcsec\ radius at z=0.05). Using the concentration
indices, we correct the stellar dispersions according to
Equation~\ref{eq_modsigma}. The results are illustrated in
Figure~\ref{bfj_10kpc}, where we compare the positions of the simulated
galaxies (red stars) with those of the FJ subset on the BFJ plane (a).
The residuals from the BFJ fit are plotted as a function of \cindx\ in
(b), and the generalized BFJ relation is shown in (c). The simulated galaxies are
all disk-dominated according to our definition
(\ie, $R_{90}/R_{50} \leq 2.8$) except one, and their stellar
dispersions are systematically offset from the fit in (a) towards
smaller values, as the disk galaxies in the GASS sample are.
The correction applied to their stellar dispersions removes the
offset, and the simulated galaxies lie on top of the relation in (c).
Although it is based on eight halos only, the excellent agreement between
simulations and data is encouraging.

\begin{figure}
\includegraphics[width=8cm]{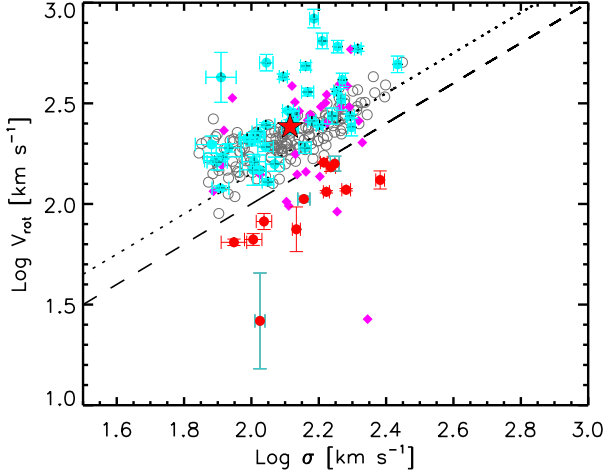}
\caption{Relation between rotational velocity and velocity dispersion
  for GASS detections. Symbols are the same as those in Figure~\ref{btf}.
  Dashed and dotted lines indicate the 1:1 correlation and the
  $V_{\rm rot} = \sqrt{2}\sigma$ relation, respectively. Error bars are
  plotted only for galaxies with inclination $i$ smaller than
  40\deg\ (cyan) or with $V_{\rm rot} < \sigma$ (red) that are not
  affected by beam confusion (magenta diamonds). Red symbols with
  cyan error bars are galaxies with both $V_{\rm rot} < \sigma$ and $i \leq 40$\deg.
}
\label{vsigma}
\end{figure}

We have shown that disk-dominated galaxies are offset in the BFJ
plane, and that this is what allow us to obtain a tight baryonic
mass-velocity relation that holds for all the galaxies in our sample.
The reason why there is a {\em disk} BFJ relation for massive galaxies is
that $\sigma$ is proportional to \vrot, and their ratio is a
function of galaxy morphology. As demonstrated by \citet{courteau07b},
for a given \vrot, earlier type galaxies have higher $\sigma$. The
brightest, bulge-dominated galaxies (ellipticals, lenticulars and
early-type spirals) lie on the $V_{\rm rot} = \sqrt{2} \sigma$ relation
expected for isothermal stellar systems. Later-type spiral and dwarf
galaxies, on the other hand, depart from the isothermal relation by an
amount that depends on morphology or total light concentration (see their Fig.~1). 
We plot the relation between \vrot\ and $\sigma$ for the \hi-detected
GASS galaxies in Figure~\ref{vsigma}. 
Our sample shows two populations of outliers, characterized by too
large or too low \vs\ ratios compared to the rest of the data set. 
These are the same outliers of the BTF relation (see Fig.~\ref{btf}); in particular,
galaxies with large \vs\ ratios all have small inclinations ($\leq
40$\deg, cyan), and objects with $V_{\rm rot} < \sigma$ (red) were
mentioned in \S~\ref{s_btf} and are discussed in the Appendix.
To avoid crowding, we show error bars for these galaxies only\footnote{
The errors on rotational velocities are obtained from standard error
propagation, taking into account the uncertainties in the observed
velocity width and inclination (\ie, $b/a$ axis ratio in Eq.~1) only.
The errors on stellar velocity dispersions are from the SDSS (as noted
in \S~\ref{s_vdisp} the aperture correction is very small, thus we do
not propagate its uncertainty).
}.
Disregarding the outliers, one can see that the galaxies with larger
stellar dispersions lie very close to the isothermal relation
(indicated by a dotted line), whereas there is a tail of galaxies with
higher \vs\ ratios at lower dispersions. The galaxies in this tail are
all disk dominated. Correcting the stellar dispersions according to 
Equation~\ref{eq_modsigma} effectively translates into accounting for this 
departure from isothermality for disk-dominated galaxies.

\subsection{Empirical Baryonic Mass - \sov\ Relation}\label{s_s05}

In the previous section, we obtained a relation
between baryonic mass and a measure of internal velocity that holds
for our sample of massive galaxies, without any morphological pruning.
We now ask whether we can find an even tighter relation by combining
\vrot\ and $\sigma$ measurements. Both quantities measure the depth of
the gravitational potential well of a galaxy, but they do so at
different radii, namely those of the bulge and of the outer disk
(as traced by the \hi\ gas, which typically extends well beyond the
stellar disk; however in very high-density 
environments \hi\ could be severely stripped; \eg, \citealt{gh85}).
In principle, \hi\ widths should provide the best measurement of the
total mass (including dark) of a galaxy, regardless of the presence of
a bulge component. However, as discussed above, reliable
\vrot\ measurements are not possible for galaxies with small inclination
to the line-of-sight --- in the limit of a perfectly
face-on system, which has no measurable rotation, the velocity
dispersion of the stellar component is a more useful quantity.
Thus, we have two methods of estimating the dynamical mass of a galaxy
that are affected by different systematics and limitations. Can we
combine them in order to obtain a quantity that is, {\it on average}, 
more tightly correlated with the baryonic mass?
An interesting quantity in this context is the \sov\ parameter
\citep[e.g.,][]{covington10}:
\begin{equation}
    S_{0.5} = \sqrt {0.5 V_{\rm rot}^2 + \sigma^2}
\label{eq_s05}
\end{equation}
This expression has been used in several studies, with
different meanings for the dispersion component. 
\citet{kassin07} studied the TF relation for an emission line-selected
sample of 544 galaxies at $0.1<z<1.2$, and demonstrated that a remarkably
tighter relation is obtained when the \sov\ parameter is adopted
instead of the rotational velocity from the optical rotation curve.
In their case, $\sigma$ is the dispersion of the gas (but this may be
contaminated by velocity gradients, see their
section 4.2). Interestingly, their sample includes early to late
spirals, irregular galaxies and merging systems, and their
\sov-stellar mass relation for the lowest redshift bin is in very good
agreement with the FJ relation of \citet{gallazzi06} in terms of slope
and zero point.

More relevant for our work, \citet{zaritsky08} used the
\sov\ parameter (which they refer to as internal velocity, $V$) with  
$\sigma$ measuring the stellar velocity dispersion of
spheroidal galaxies. They showed
that all classes of galaxies lie on a two-dimensional surface,
which is expressed as a linear combination of the logarithms of the
effective radius $r_{\rm e}$, the internal velocity squared, the surface brightness
within $r_{\rm e}$, and the mass-to-light ratio within $r_{\rm e}$. Their sample
is a heterogeneous collection of 1925 spheroids and disk galaxies from
existing data sets, which span the full range of galaxy types and
luminosities. However, although they define $V^2 \equiv 0.5 V_{\rm c}^2 +\sigma^2$,
they use {\it either} the circular velocity $V_{\rm c}$ for disk galaxies or
the stellar velocity dispersion $\sigma$ for spheroids, but never add
the two contributions. 

Another interesting analysis was carried out by \citet{covington10},
who studied the evolution of the stellar mass TF relation in a
simulation of a disk merger. They argue that, since early-type
galaxies are generally assumed to form through mergers of late-type
systems, the scaling relations of early types should descend from
those of late-type galaxies. Thus, they simulate the evolution of a
galaxy merger, mimic observations of emission lines, and study
how the kinematics of the system changes with time. The progenitors
are two identical Sbc galaxies lying on the stellar mass TF relation,
which merge and form a rotating elliptical galaxy. Intriguingly, they
show that, while rotation is converted into stellar dispersion, the
\sov\ parameter is approximately conserved --- suggesting that
\sov\ might be really tracing the mass distribution.

\begin{figure*}
\includegraphics[width=17cm]{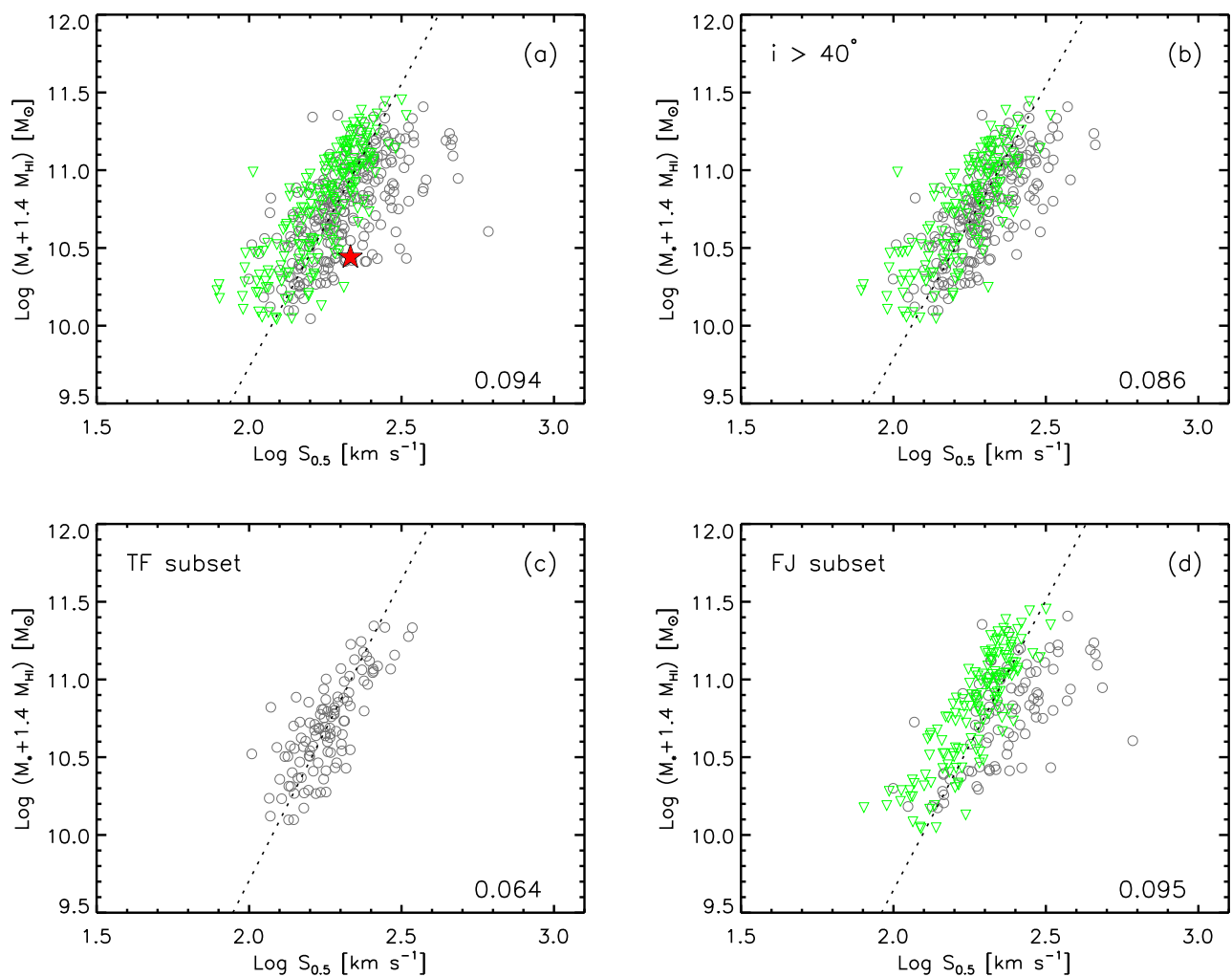}
\caption{\sov\ relations for the full sample (a), for the subset of galaxies with
  inclination larger than 40\deg\ (b), and for the TF (c) and FJ (d) subsets.
  \hi\ non-detections are indicated as green triangles; the red
  star in (a) corresponds to GASS 3505. Dotted lines
  are inverse fits to the data points (the scatter is noted in the
  bottom right corner of each panel).}
\label{s05}
\end{figure*}

The above studies motivated us to look at the relation between baryonic
mass and \sov\ parameter for the galaxies in our sample, with an
important caveat. The \sov\ parameter defined by equation~\ref{eq_s05}
has a physical meaning only if rotation and dispersion are associated to
the {\it same} system. For instance, for a rotating isothermal sphere
the dynamical support is provided by both ordered and thermal motions,
and the above combination is a direct result of the virial theorem. 
For our galaxies, \vrot\ measures the rotational velocity of the gas, and
$\sigma$ is the velocity dispersion of the stars in the bulge, thus
combining the two {\em has no physical motivation}. However, we
are not trying to partition the dynamical support of a galaxy between
its bulge and disk components --- as we already pointed out, \vrot\ should
account for the full dynamical support, regardless of the presence of
a bulge. We wish instead to establish if we can average two measures
of the potential well of a galaxy, which are based on different
tracers, in order to obtain a quantity that, empirically, better
correlates with the baryonic mass. After all, we expect the \hi\ width
to be a more reliable measurement for inclined disk galaxies, and
$\sigma$ to be better for elliptical or face-on, disk galaxies. Thus,
the combination of the two might work better {\em on average}.

The correlation between baryonic mass and \sov\ parameter is
plotted in Figure~\ref{s05} for the full sample (a), for the
galaxies with inclinations larger than 40\deg\ (b), and for the TF and
FJ subsets (c, d) defined in \S~\ref{s_btf} and \S~\ref{s_bfj}.
The \sov\ relation for the full sample has a scatter of 0.094 dex,
very close to that of the BFJ relation for the full sample, but significantly
larger than that of the generalized BFJ plotted in Figure~\ref{bfj_corr}.
The \sov\ relation still suffers from the
inclination problems that affect the BTF, although to a
lesser degree. The strongest outliers seen in Figure~\ref{s05}a are removed by our
40\deg\ inclination cut (b,c). Notice that, when applied to
disk-dominated galaxies with inclinations larger than 40\deg, the
\sov\ relation has a scatter of only 0.064 dex, \ie, it is tighter than the BTF shown in
Figure~\ref{btf}b (which has a scatter of 0.076 dex) --- this is in
fact the tightest relation presented in this work (for comparison, the
generalized BFJ relation restricted to the same sample has a scatter of 0.077 dex). 
Conversely, restricting the sample to bulge-dominated galaxies does
not decrease the scatter of the \sov\ relation. There is still a small
offset between galaxies with and without \hi\ detections. Naturally,
we do not have any information on the rotational velocity of the
non-detections (some of which are disks close to edge-on view) 
--- thus, for a fraction of these, \sov\ is
certainly underestimated. 
Overall, the baryonic \sov\ relation is as good as our generalized BFJ, but only for
galaxies with inclinations larger than 40\deg.

\begin{figure*}
\includegraphics[width=17cm]{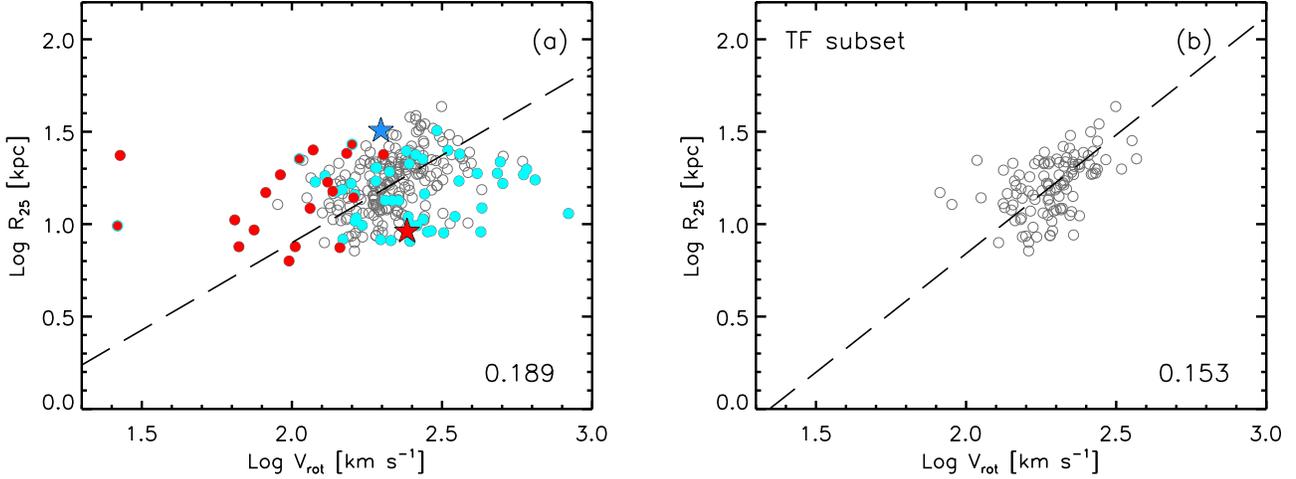}
\caption{Relation between optical size and rotational
  velocity for all the galaxies with \hi\ detections (a) and for the
  TF subset (b). Symbols are the same as those in Figure~\ref{btf}.
  Dashed lines are bisector fits to the data points (the scatter is
  noted in the bottom right corner).}
\label{vsize}
\end{figure*}

\section{Size-Velocity Relations}\label{s_vsize}

In this work we demonstrated that a simple correction applied
to the stellar velocity dispersions removes the offset between 
disk-dominated galaxies and spheroids seen in the BFJ relation,
\ie, to first order, it removes its dependence
on the internal structure of the galaxy.
We further investigate this by considering the size-velocity
relation.  The existence of such a
correlation is well established for both disk galaxies 
(\eg, \citealt{courteau07a,avila-reese08}; SS11 and references therein)
and ellipticals or bulges of early-type spirals
(the $D_{\rm n}$-$\sigma$ relation, where $D_{\rm n}$ is the diameter of
the galaxy at a given surface brightness level; \eg,
\citealt{dressler87a,bernardi02}). Here, we compare this relation for
the different velocity indicators and data subsets discussed in this work.
Specifically, we wish to test how the correction to the stellar
velocity dispersion introduced in \S~\ref{s_mbfj}
affects the size-velocity relation.

The sizes of disk galaxies are commonly estimated from exponential
scale lengths, \rd. However, disk scale lengths are not only problematic
to measure (\eg\ \citealt{giovanelli94}; SS11), but are
also not meaningful for elliptical galaxies. \citet{saintonge08} and SS11
advocate the use of isophotal radii instead of disk scale lengths, as
they yield scaling relations with significantly lower
scatter. Isophotal radii are also used as size indicators for
early-type galaxies in the $D_{\rm n}$-$\sigma$ relation.
Thus, we adopt $R_{25}$, which is half the 25 mag arcsec$^{-2}$
isophote diameter $D_{25}$ measured by us on the SDSS {\em g}-band
images, as a measure of size for all the galaxies in our sample.
For comparison, $R_{25}$ varies between 2 and 7 \rd, where
\rd\ is the SDSS exponential scale length in \rband, for the
disk-dominated galaxies in our sample.

The relation between size and rotational velocity (RV) is shown in
Figure~\ref{vsize}a for all the galaxies with \hi\ detections. 
For this and the other relations presented in this section we
performed bisector linear regressions instead of inverse fits, because
it is less clear that the scatter is mostly confined to one variable. 
The scatter of $R_{25}$ about the best fit is computed as before, \ie,
by applying Tukey's bi-weight, and is noted in the bottom right corner of
the plot (in dex of kpc).
As in Figures~\ref{btf} and \ref{vsigma}, cyan and red circles
indicate galaxies with inclinations smaller than 40\deg\ and with
$V_{\rm rot}/\sigma < 1$,
respectively. Some of the main outliers of this relation are also
outliers for the BTF, but there is not a 1-to-1 correspondence.
For instance, the very \hi-rich spiral GASS 35981
(marked as a blue star) is an outlier on this plot, but not on the
BTF, whereas the gas-rich elliptical GASS 3505 (red star) is an
outlier for both relations.

\begin{figure*}
\includegraphics[width=17cm]{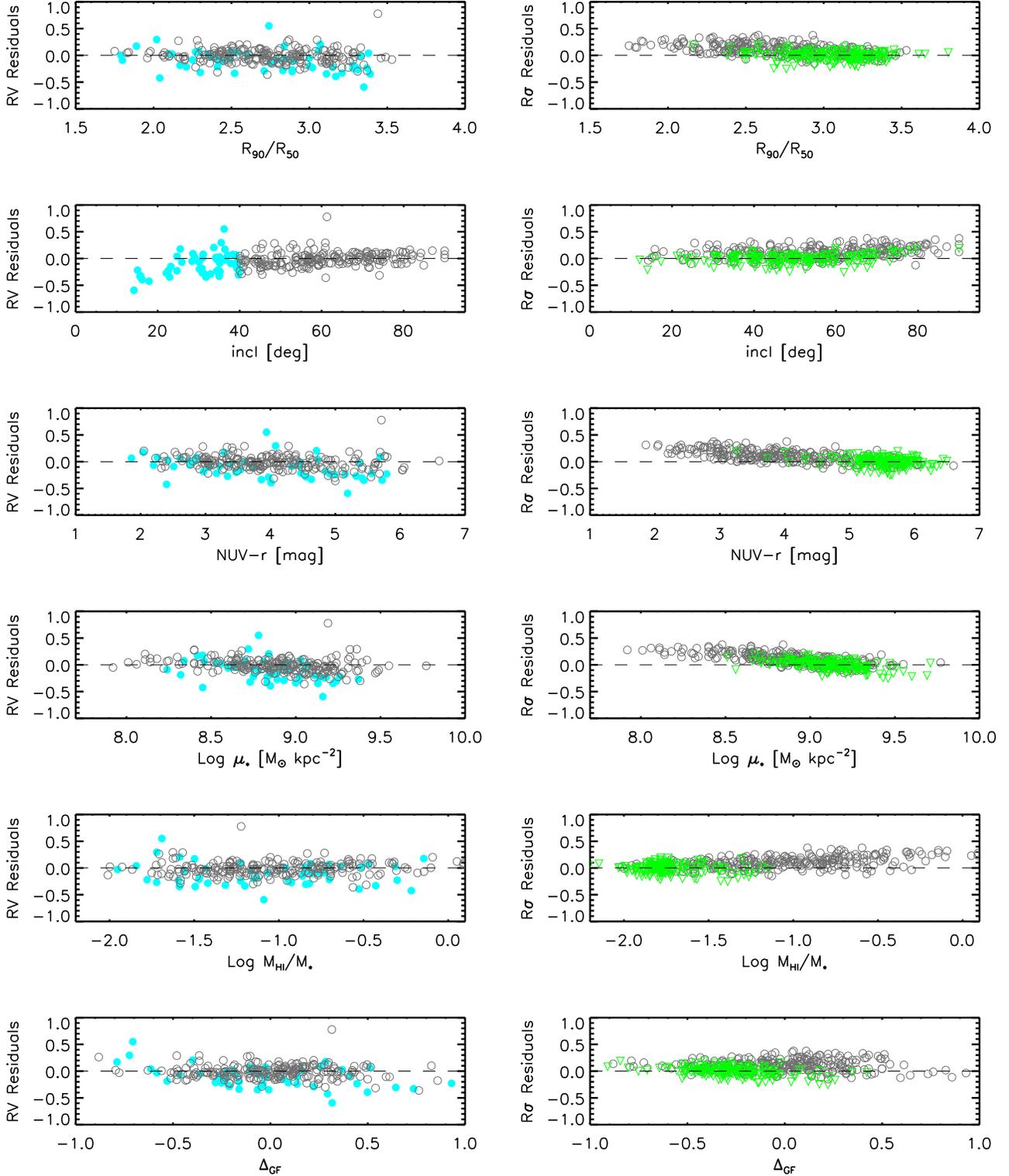}
\caption{Residuals of RV (left column) and R$\sigma$ (right column) 
  relations plotted as functions of concentration index
  (row 1), galaxy inclination (row 2), \nuvr\ color (row 3), stellar
  mass surface density (row 4),
  gas fraction (row 5), and distance from the gas fraction plane
  (row 6). The residuals are computed from the orthogonal distances to
  the bisector fits shown in Figures~\ref{vsize}b and
  \ref{sigma_size}b, respectively. Cyan and green symbols indicate
  galaxies with inclinations smaller than 40\deg\ and \hi\ non-detections, respectively.}
\label{vsize_res}
\end{figure*}

The RV relation is significantly weaker and more scattered than the
BTF, as known from other studies that use disk scale lengths as size
indicators (\eg, \citealt{courteau07a} report a Pearson correlation
coefficient of $r \sim 0.65$ and a scatter of 0.33 in log $R$).
However, as mentioned above, SS11 obtain
a very tight correlation using isophotal radii instead of
disk scale lengths ($r=0.84$ and a scatter of 0.11 in log R; in
the notation adopted by these authors, the scatter of Courteau's
relation becomes 0.165 dex).
When restricted to the TF subset (Fig.~\ref{vsize}b), the scatter
of our relation (0.15 dex) is intermediate between those of 
\citet{courteau07a} and SS11 samples.
Although we use a similar size indicator as SS11,
their sample includes only late-type spirals, thus the larger
scatter of our relation is most likely due to the broader morphological mix
of the galaxies in our data set.
As for the BTF and BFJ relations, we plot the residuals of the RV
relation as a function of several quantities in
Figure~\ref{vsize_res} (left panels). There is a dependence of the RV
residuals on inclination and, to a lesser extent, on \must.

\begin{figure*}
\includegraphics[width=17cm]{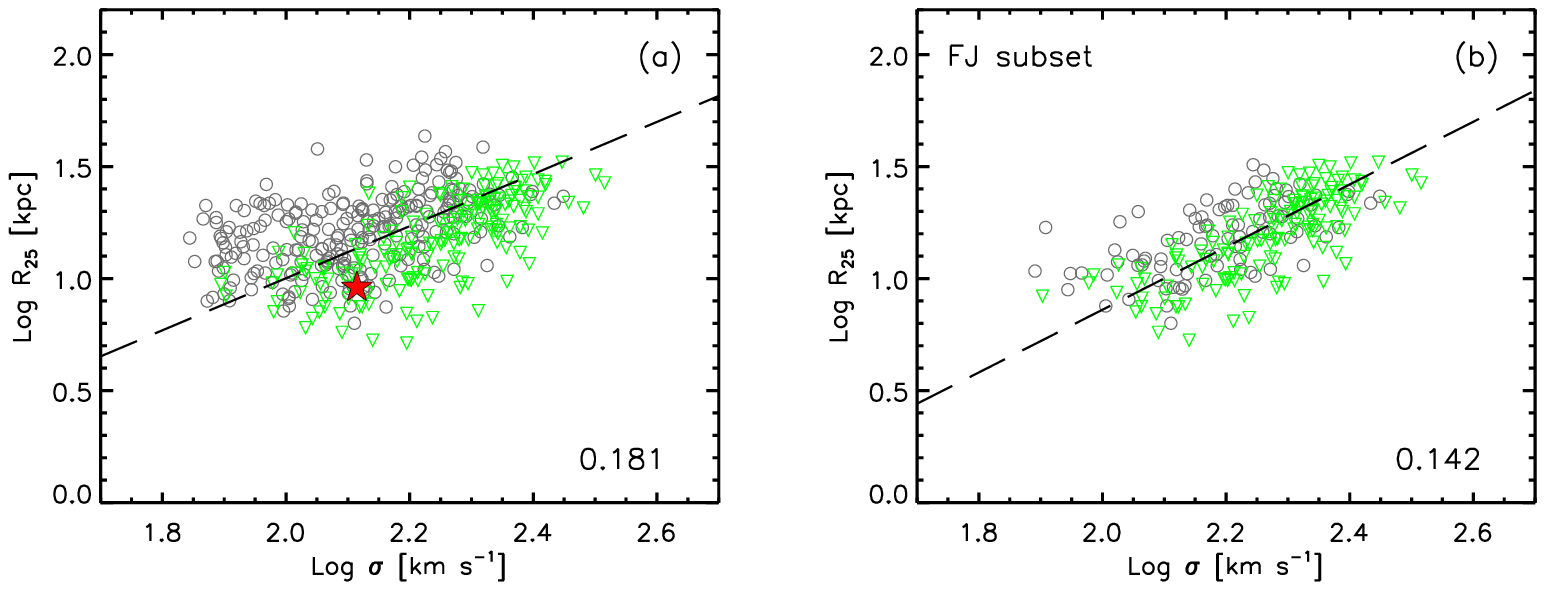}
\caption{Relation between optical size and stellar velocity
  dispersion for the full sample (a) and for the
  FJ subset (b). Symbols are the same as those in Figure~\ref{s05}.
  Dashed lines are bisector fits to the data points (the scatter is
  noted in the bottom right corner).}
\label{sigma_size}
\end{figure*}

Figure~\ref{sigma_size} shows the relation between $R_{25}$ and stellar
velocity dispersion $\sigma$ (R$\sigma$) for the full sample (a) and for the
FJ subset (b). As for the BFJ relation, galaxies with \hi\ detections are
displaced from the non-detections (green triangles), and a
clear trend is observed when the R$\sigma$ residuals 
are plotted as a function of concentration index
(Fig.~\ref{vsize_res}, top right panel).
Indeed, Figure~\ref{vsize_res} shows that the R$\sigma$ residuals
behave similarly to the BFJ ones (see Fig.~\ref{residuals}):
disk-dominated galaxies systematically deviate from the best fit
relation obtained for the FJ subset, which is dominated by spheroids.
Aside from systematic trends, which are stronger for the R$\sigma$
relation, the scatter of the RV and R$\sigma$ relations is very
similar, for both maximum samples and pruned subsets.

\begin{figure}
\includegraphics[width=8cm]{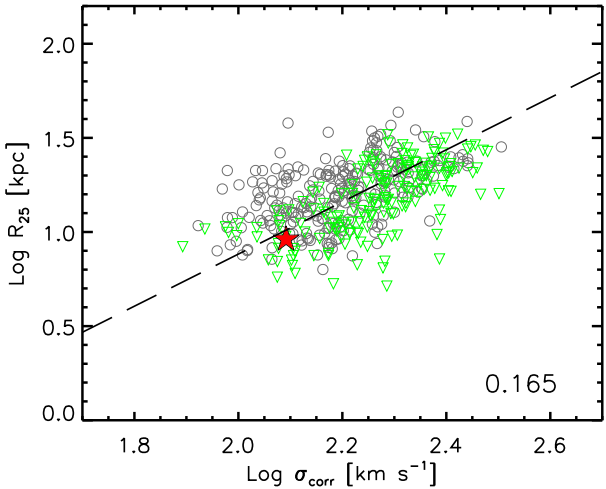}
\caption{Same as Figure~\ref{sigma_size}a for the corrected velocity dispersion.}
\label{sigmacorr_size}
\end{figure}

Lastly, correcting the velocity dispersions according to 
Equation~\ref{eq_modsigma} yields a slightly tighter size-velocity relation 
(Fig.~\ref{sigmacorr_size}). Most importantly, our correction removes
the offset between \hi\ detections and non-detections present in 
Figure~\ref{sigma_size}a (the remaining few \hi\ detections lying
above the relation are highly inclined galaxies, for which 
$R_{25}$, which is not corrected for inclination, is likely to be
overestimated).

\section{Discussion and Conclusions}\label{s_disc}

The main result of this work is the existence of a tight relation
between baryonic mass and velocity that is independent of galaxy
morphology. To first order, we can remove the dependence of
dynamical scaling relations on the internal structure of the galaxy
by applying a simple correction to the measured stellar velocity
dispersions, which depends only on the concentration index \cindx\
(Eq.~\ref{eq_modsigma}). The correlation between baryonic mass and
corrected $\sigma$ thus obtained has a scatter of only 0.08 dex
(Fig.~\ref{bfj_corr} and Table~\ref{t_fits}). 
It is encouraging that our correction removes the offset between disks
and spheroids also in the size-velocity relation (expressed in terms
of $R_{25}$ versus stellar dispersion).
We tested if an even tighter baryonic mass-velocity relation could be obtained by using the
\sov\ parameter, which combines rotational velocity and stellar
dispersion. We found no improvement, because the \sov\ relation still
suffers from the problem of low inclination galaxies that affects
the BTF, to a lesser extent. Despite being measured at a smaller
spatial scale than the \hi\ width, the stellar velocity dispersion
turns out to be a better tracer of mass, at least for the massive
galaxies in our sample.
This is surprising, as one would expect the rotational velocity
to provide a more reliable measurement of mass for galaxies with
\hi\ detections, the vast majority of which are rotation dominated
(\ie, 216/228 galaxies have $V_{\rm rot}/\sigma >1$), regardless of the
presence of a bulge.
And yet, the same result was obtained by \citet{eyal99} for a sample
of S0 galaxies with inclinations \about 35\deg $-$60\deg, 
which are also, overall, rotation-dominated: the
central stellar velocity dispersion is a better predictor of
\iband\ luminosity than the circular speed at 2-3 exponential disk
scale lengths, which they carefully measured from long-slit optical
absorption-line spectra.

As we already pointed out, the main limitation with rotational
velocities is observational, not intrinsic. We are simply not able to
reliably measure circular speeds from line-of-sight velocities, except
for well-selected samples of undisturbed late-type, inclined spirals
(\ie, the typical TF samples), for which the deprojection to edge-on
view does not introduce very large uncertainties.  As for the velocity
dispersions, which are measured through 3\arcsec -diameter fibers,
there might be a concern about contamination from disk stars in
circular orbits, especially for the most edge-on, disk-dominated
galaxies. However, we have argued that this effect is
negligible (\S~\ref{s_bfj}). This agrees with the conclusion reached
by \citet{courteau07b},  who reported that the contamination is small
($<5\%$, based on simulations) for Milky Way-type galaxies with
pressure-supported bulges.

We argued that the reason why disk-dominated galaxies are simply
offset from the spheroids on the BFJ plane (which is the key point
that allows us to bring all massive galaxies onto the same baryonic
mass-velocity relation) is that $\sigma$ is proportional
to \vrot, and their ratio is a function of galaxy morphology.
Although initial work indicated that spirals and ellipticals follow
the same tight \vrot-$\sigma$ correlation for $\sigma > 80$ \kms\
\citep[\eg,][]{ferrarese02,baes03,pizzella05},
more recent analyses demonstrate that the relation between
\vrot\ and $\sigma$ is not universal, but depends on morphology.
As mentioned in \S~\ref{s_mbfj}, \citet{courteau07b} show that the brightest,
bulge-dominated galaxies lie on the $V_{\rm rot} = \sqrt{2} \sigma$
relation expected for isothermal stellar systems, whereas 
later-type spirals and dwarfs are offset by an amount that depends on
morphology or total light concentration. We have shown that correcting
the stellar dispersions according to Equation~\ref{eq_modsigma}
effectively translates into accounting for this departure from
isothermality for disk-dominated galaxies.
\citet{courteau07b} noted that, despite the fact that a
detailed understanding of what sets the relation between \vrot,
$\sigma$ and concentration index is still missing, one could use this
relation to empirically reduce the scatter of scaling relations that
involve dynamical parameters, such as the TF or FJ. We showed that
the existence of such relation allows us to do more than that --- we
obtained a generalized BFJ relation that holds for all the massive galaxies in
our sample, with a scatter (0.079 dex) that is as small as that of the
BTF and BFJ relations applied to their pruned subsets (\ie\ 0.076 and
0.074 dex, respectively). For comparison, \citet{avila-reese08} and 
\citet{gallazzi06} report a scatter of 0.06 dex for the BTF and 0.071
dex for the stellar FJ relations, respectively.

The implications of our generalized BFJ relation for extragalactic studies are very
promising. Because it holds for all massive galaxies in our sample
{\em regardless of morphology}, this relation appears to provide a
more fundamental link between dark matter halo mass and baryonic
content than that obtained using rotational velocities or velocity
dispersions. As such, it gives more fundamental constraints to galaxy
formation models than the TF or FJ/FP relations.
Also this is the reference baryonic mass-internal velocity
relation that higher redshift studies, which naturally target the most
massive galaxies, should compare with. This relation is more resilient
to systematic effects than the TF, and, contrary to both TF and FJ/FP
relations, does not require any morphological pruning --- a
significant advantage since accurate morphological classifications are
difficult to obtain for large samples beyond the very local Universe.

As pointed out throughout this work, our results are based on a
representative sample of galaxies with stellar masses larger than
$10^{10}$ \Msun. It remains to be established how far down in stellar
mass these results can be extrapolated. It would be beneficial to 
investigate whether a BFJ relation for disk galaxies holds down to 
low baryonic masses (where the gas contribution is more important),
and with similarly low scatter, based on a representative and
homogeneous sample such as GASS.

\section*{Acknowledgments}

B.C. wishes to thank Dennis Zaritsky, Susan Kassin, Thorsten Naab
and Luca Cortese for useful discussions. Thanks also to Simone
Weinmann for kindly providing the data on concentration index versus
B/T ratio used in section 3.2.

The Arecibo Observatory is part of the National Astronomy and
Ionosphere Center, which is operated by Cornell University under a
cooperative agreement with the National Science Foundation. 

GALEX (Galaxy Evolution Explorer) is a NASA Small Explorer, launched
in April 2003. We gratefully acknowledge NASA's support for
construction, operation, and science analysis for the GALEX mission,
developed in cooperation with the Centre National d'Etudes Spatiales
(CNES) of France and the Korean Ministry of Science and Technology. 

Funding for the SDSS and SDSS-II has been provided by the Alfred
P. Sloan Foundation, the Participating Institutions, the National
Science Foundation, the U.S. Department of Energy, the National
Aeronautics and Space Administration, the Japanese Monbukagakusho, the
Max Planck Society, and the Higher Education Funding Council for
England. The SDSS Web Site is http://www.sdss.org/.

The SDSS is managed by the Astrophysical Research Consortium for the
Participating Institutions. The Participating Institutions are the
American Museum of Natural History, Astrophysical Institute Potsdam,
University of Basel, University of Cambridge, Case Western Reserve
University, University of Chicago, Drexel University, Fermilab, the
Institute for Advanced Study, the Japan Participation Group, Johns
Hopkins University, the Joint Institute for Nuclear Astrophysics, the
Kavli Institute for Particle Astrophysics and Cosmology, the Korean
Scientist Group, the Chinese Academy of Sciences (LAMOST), Los Alamos
National Laboratory, the Max-Planck-Institute for Astronomy (MPIA),
the Max-Planck-Institute for Astrophysics (MPA), New Mexico State
University, Ohio State University, University of Pittsburgh,
University of Portsmouth, Princeton University, the United States
Naval Observatory, and the University of Washington.

\begin{figure*}
\includegraphics[width=17cm]{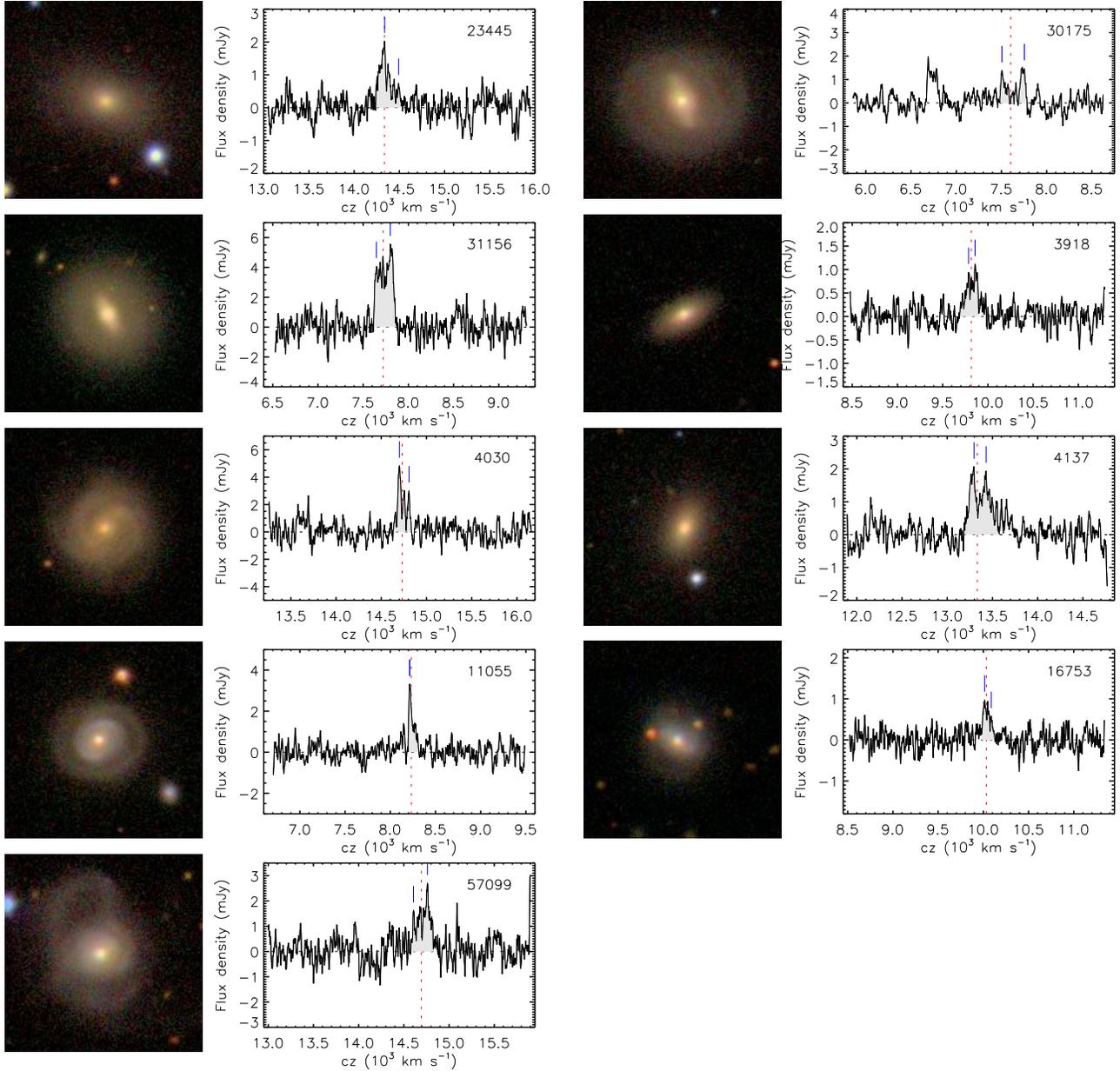}
\caption{SDSS postage stamps (1\arcmin\ size) and \hi\ spectra of subset of outliers
  marked in red in Figures~\ref{btf} and \ref{vsigma}.
  The GASS identifier of each galaxy is indicated on the top right
  corner of its spectrum. A dotted line and two
  dashes indicate the heliocentric velocity corresponding to the SDSS
  redshift and the two peaks used for width measurement, respectively.
  First row and GASS 31156: GASS DR1 (Paper 1); other objects: GASS
  DR2 (Catinella et al., in preparation).
}
\label{vsigma_hi}
\end{figure*}

\section*{Appendix: Galaxies with  $V_{\rm rot} < \sigma$}

We discuss here in more detail the 12 galaxies with \hi\ rotational
velocities smaller than their stellar velocity dispersions, which are
outliers of both BTF and \vrot\ versus $\sigma$ relations.
These objects are indicated by red symbols in 
Figures~\ref{btf} and \ref{vsigma}. 
While the high \vs\ outliers are all galaxies with small
inclinations, for which the rotational velocities are likely to be
overestimated, the low \vs\ outliers are potentially more interesting.
As mentioned in \S~\ref{s_btf}, \citet{ho07} identifies a population of
galaxies with unusually small \vs\ ratios, which are outliers in his
TF relation, and argues that these systems must have experienced gas
accretion. 
Figure~\ref{vsigma_hi} shows SDSS images and
\hi -line profiles for 9 outliers detected by GASS (the other
galaxies were either observed by ALFALFA or included in the S05
archive). 
Two of these galaxies, GASS 30175 and 31156, are outliers because
their SDSS inclinations are incorrect. They both have strong bars
which dominate the SDSS fits, thus overestimating the inclination
(54\deg\ and 61\deg\ for GASS 30175 and 31156; we measured 37\deg\ and
27\deg\ respectively). The same is true for GASS 24236 (AGC 220363,
detected by ALFALFA and not shown here), a nearly face-on barred
galaxy with SDSS inclination of 46\deg\ (we measured 17\deg).
Of the remaining outliers, one is clearly morphologically disturbed
(GASS 57099, bottom row). 
Another galaxy, GASS 4030, was observed as part of the COLDGASS
survey \citep[][see their Figure A1, row 6]{coldgass1}. Its
(uncorrected) velocity width measured from the CO(1-0) spectrum is 
137 $\pm 11$ \kms, which is exactly the same as the (uncorrected)
\hi\ width. Since the CO(1-0) emission typically originates from the
central regions of a galaxy, perhaps the fact that CO and \hi\ widths
are identical means that the \hi\ gas in this galaxy is not extended
enough to sample the flat part of the rotation curve. 
This is not particularly unusual, as there are other examples of 
early-type spirals with truncated \hi\ disks (\eg\ NGC 3623,
where the \hi\ emission does not extend beyond the
stellar disk; \citealt{hogg01}) or unexpectedly narrow \hi\ profiles
(\eg\ NGC 5854; \citealt{haynes00}).
Overall, even acknowledging that some of these profiles are not very 
high signal-to-noise detections, it is striking that most are
asymmetric, suggesting disturbances in the distribution
and/or kinematics of the \hi\ gas. We will investigate asymmetries of
\hi\ profiles in more detail in a future work.

\begin{table*}
\centering
\caption{Fits to Baryonic Mass - Velocity Relations}
\label{t_fits}
\begin{tabular}{lccccc}
\hline
                        &                  &  &  \multicolumn{2}{c}{Log $y$= a Log $x$+b$^{[b]}$}  &   \\
                        &  $y$             &  N$^{[a]}$ &  a   &   b                &  scatter$^{[c]}$\\
\hline
BTF, \hi\ detections	&  \vrot	   &  259  &  0.250  &   $-$0.365   &  0.127  \\
BTF, TF subset	        &  \vrot	   &  108  &  0.237  &   $-$0.251   &  0.076  \\
BFJ                 	&  $\sigma$	   &  436  &  0.299  &   $-$1.053   &  0.107  \\  
BFJ, FJ subset	        &  $\sigma$	   &  231  &  0.283  &   $-$0.818   &  0.074  \\  
Generalized BFJ       	&$\sigma_{\rm corr}$&  436  &  0.267  &   $-$0.654   &  0.079  \\  
Baryonic \sov           &  \sov	           &  436  &  0.274  &   $-$0.670   &  0.094  \\ 
Baryonic \sov, $i>40$\deg  &  \sov	   &  341  &  0.285  &   $-$0.791   &  0.086  \\ 
\hline
\end{tabular}
\begin{flushleft}
$^{[a]}$Number of galaxies in the sample.\\
$^{[b]}$Inverse fits to the data points shown in Figures~\ref{btf},
  \ref{bfj}, \ref{bfj_corr}, and \ref{s05}; $x\equiv \rm M_\star+1.4~M_{HI}$.\\
$^{[c]}$Scatter in dex of \kms.
\end{flushleft}
\end{table*}

\begin{table*}
\centering
\caption{Fits to Size - Velocity Relations}
\label{t2_fits}
\begin{tabular}{lccccc}
\hline
                        &                  &  & \multicolumn{2}{c}{Log $R_{25}$= a Log $x$+b$^{[b]}$} &\\
                        &  $x$             &  N$^{[a]}$ &  a   &   b                   &  scatter$^{[c]}$\\
\hline
RV, \hi\ detections	&  \vrot	   &  259  &  0.946  &   $-$0.992   &  0.189  \\
RV, TF subset	        &  \vrot	   &  108  &  1.284  &   $-$1.728   &  0.153  \\
R$\sigma$               &  $\sigma$	   &  436  &  1.162  &   $-$1.323   &  0.181  \\  
R$\sigma$, FJ subset	&  $\sigma$	   &  231  &  1.399  &   $-$1.937   &  0.142  \\  
R$\sigma_{\rm corr}$  	&  $\sigma_{\rm corr}$ &  436  &  1.385  &   $-$1.887   &  0.165  \\  
\hline
\end{tabular}
\begin{flushleft}
$^{[a]}$Number of galaxies in the sample.\\
$^{[b]}$Bisector fits to the data points shown in Figures~\ref{vsize},
  \ref{sigma_size}, and \ref{sigmacorr_size}.\\
$^{[c]}$Scatter in dex of kpc.\\
\end{flushleft}
\end{table*}


\begin{thebibliography}{}

\bibitem[\protect\citeauthoryear{Abazajian et al.}{2009}]{sdss7} 
        Abazajian, K. N. et al. 2009, ApJS, 182, 543

\bibitem[\protect\citeauthoryear{Avila-Reese et al.}{2008}]{avila-reese08}
        Avila-Reese, V., Zavala, J., Firmani, C., \& Hern{\'a}ndez-Toledo,
        H. M. 2008, AJ, 136, 1340

\bibitem[\protect\citeauthoryear{Baes et al.}{2003}]{baes03} 
        Baes, M., Buyle, P., Hau, G. K. T., \& Dejonghe, H. 2003, MNRAS, 341, L44

\bibitem[\protect\citeauthoryear{Baldry et al.}{2006}]{baldry06} 
        Baldry, I. K., Balogh, M. L., Bower, R. G., Glazebrook, K.,
        Nichol, R. C., Bamford, S. P., \& Budavari, T. 2006, MNRAS, 373, 469

\bibitem[\protect\citeauthoryear{Bedregal et al.}{2006}]{bedregal06} 
        Bedregal, A. G., Arag{\'o}n-Salamanca, A., \& Merrifield,
        M. R. 2006, MNRAS, 373, 1125

\bibitem[\protect\citeauthoryear{Begum et al.}{2008}]{begum08} 
        Begum, A., Chengalur, J. N., Karachentsev, I. D., \& Sharina, M. E.
        2008, MNRAS, 386, 138

\bibitem[\protect\citeauthoryear{Bernardi et al.}{2002}]{bernardi02} 
        Bernardi, M., Alonso, M. V., da Costa, L. N., Willmer, C. N. A., Wegner, G.,
        Pellegrini, P. S., Rit{\'e}, C., \& Maia, M. A. G. 2002, AJ, 123, 2159

\bibitem[\protect\citeauthoryear{Bernardi et al.}{2003}]{bernardi03} 
        Bernardi, M. et al. 2003, AJ, 125, 1817

\bibitem[\protect\citeauthoryear{Bruzual \& Charlot}{2003}]{bruzual03} 
       	Bruzual, G., \& Charlot, S. 2003, MNRAS, 344, 1000

\bibitem[\protect\citeauthoryear{Catinella et al.}{2006}]{templates} 
        Catinella, B., Giovanelli, R., \& Haynes, M. P. 2006, ApJ, 640, 751

\bibitem[\protect\citeauthoryear{Catinella et al.}{2007}]{widths} 
        Catinella, B., Haynes, M. P., \& Giovanelli, R. 2007, AJ, 134, 334

\bibitem[\protect\citeauthoryear{Catinella et al.}{2008}]{highz} 
        Catinella, B., Haynes, M. P., Giovanelli, R., Gardner, J. P., 
        \& Connolly, A. J. 2008, ApJ, 685, L13

\bibitem[\protect\citeauthoryear{Catinella et al.}{2010}]{gass1} 
        Catinella, B., Schiminovich, D., Kauffmann, G. et al. 2010,
        MNRAS, 403, 683 (Paper I)

\bibitem[\protect\citeauthoryear{Chabrier}{2003}]{chabrier03} 
        Chabrier, G. 2003, PASP, 115, 763 

\bibitem[\protect\citeauthoryear{Courteau}{1997}]{courteau97}
        Courteau, S. 1997, AJ, 114, 2402

\bibitem[\protect\citeauthoryear{Courteau et al.}{2003}]{courteau03}
        Courteau, S., Andersen, D. R., Bershady, M. A., MacArthur,
        L. A., \& Rix, H.-W. 2003, ApJ, 594, 208

\bibitem[\protect\citeauthoryear{Courteau et al.}{2007a}]{courteau07a} 
        Courteau, S., Dutton, A. A., van den Bosch, F. C., MacArthur, L. A.,
        Dekel, A., McIntosh, D. H., \& Dale, D. A. 2007a, ApJ, 671, 203

\bibitem[\protect\citeauthoryear{Courteau et al.}{2007b}]{courteau07b} 
        Courteau, S., McDonald, M., Widrow, L. M., \& Holtzman, J. 2007b, ApJ, 655, L21

\bibitem[\protect\citeauthoryear{Courteau \& Rix}{1999}]{courteau99}
        Courteau, S. \& Rix, H.-W. 1999, ApJ, 513, 561

\bibitem[\protect\citeauthoryear{Courtois et al.}{2009}]{courtois09} 
        Courtois, H. M., Tully, R. B., Fisher, J. R., Bonhomme, N., Zavodny, M., \&
        Barnes, A. 2009, AJ, 138, 1938

\bibitem[\protect\citeauthoryear{Covington et al.}{2010}]{covington10}
        Covington, M. D. et al. 2010, ApJ, 710, 279

\bibitem[\protect\citeauthoryear{De Rijcke et al.}{2007}]{derijcke07}
        De Rijcke, S., Zeilinger, W. W., Hau G. K. T., Prugniel, P.,
        \& Dejonghe H., 2007, ApJ, 659, 1172

\bibitem[\protect\citeauthoryear{Djorgovski \& Davis}{1987}]{dd87} 
        Djorgovski, S. \& Davis, M. 1987, ApJ, 313, 59

\bibitem[\protect\citeauthoryear{Dressler et al.}{1987}]{dressler87a} 
        Dressler, A. 1987, ApJ, 317, 1

\bibitem[\protect\citeauthoryear{Dressler et al.}{1987}]{dressler87b} 
        Dressler, A., Lynden-Bell, D., Burstein, D., Davies, R. L.,
        Faber, S. M., Terlevich, R., \& Wegner, G. 1987, ApJ, 313, 42

\bibitem[\protect\citeauthoryear{Driver et al.}{2007}]{driver07} 
       	Driver, S. P., Allen, P. D., Liske, J., \& Graham, A. W. 2007,
        ApJ, 657, L85

\bibitem[\protect\citeauthoryear{Dutton et al.}{2007}]{dutton07} 
        Dutton, A. A., van den Bosch, F. C., Dekel, A., \& Courteau,
        S. 2007, ApJ, 654, 27

\bibitem[\protect\citeauthoryear{Faber \& Jackson}{1976}]{fj76} 
        Faber, S. M. \& Jackson, R. E. 1976, ApJ, 204, 668

\bibitem[\protect\citeauthoryear{Ferrarese}{2002}]{ferrarese02} 
        Ferrarese, L. 2002, ApJ, 578, 90

\bibitem[\protect\citeauthoryear{Gadotti}{2009}]{gadotti09} 
        Gadotti, D. A. 2009, MNRAS, 393, 1531

\bibitem[\protect\citeauthoryear{Gadotti \& Kauffmann}{2009}]{gadottikauffmann09} 
        Gadotti, D. A. \& Kauffmann, G. 2009, MNRAS, 399, 621

\bibitem[\protect\citeauthoryear{Gallazzi et al.}{2005}]{gallazzi05} 
        Gallazzi, A., Charlot, S., Brinchmann, J., White, S. D. M. \&
        Tremonti, C. A. 2005, MNRAS, 362, 41

\bibitem[\protect\citeauthoryear{Gallazzi et al.}{2006}]{gallazzi06} 
        Gallazzi, A., Charlot, S., Brinchmann, J., \& White, S. D. M. 2006, MNRAS, 370, 1106

\bibitem[\protect\citeauthoryear{Geha et al.}{2006}]{geha06} 
        Geha, M., Blanton, M., Masjedi, M., \& West, A. 2006, ApJ, 653, 240

\bibitem[\protect\citeauthoryear{Giovanelli et al.}{1997a}]{giovanelli97a}  
	Giovanelli, R., Haynes, M. P., Herter, T., Vogt, N. P., Wegner, G.
        Salzer, J. J., da Costa, L. N., \& Freudling, W. 1997a, AJ, 113, 22

\bibitem[\protect\citeauthoryear{Giovanelli et al.}{1997b}]{giovanelli97b}  
	Giovanelli, R., Haynes, M. P., Herter, T., Vogt, N. P., da Costa, L. N., 
        Freudling, W., Salzer, J. J., \& Wegner, G. 1997b, AJ, 113, 53

\bibitem[\protect\citeauthoryear{Giovanelli et al.}{1994}]{giovanelli94}  
	Giovanelli, R., Haynes, M. P., Salzer, J. J., Wegner, G., da
        Costa, L. N., \& Freudling, W. 1994, AJ, 107, 2036

\bibitem[\protect\citeauthoryear{Giovanelli et al.}{2005}]{alfalfa}  
	Giovanelli, R. et al. 2005, AJ, 130, 2598

\bibitem[\protect\citeauthoryear{Giovanelli \& Haynes}{1985}]{gh85}  
	Giovanelli, R. \& Haynes, M. P. 1985, ApJ, 292, 404

\bibitem[\protect\citeauthoryear{Graves et al.}{2009}]{graves09} 
        Graves, G. J., Faber, S. M., \& Schiavon, R. P. 2009, ApJ, 693, 486

\bibitem[\protect\citeauthoryear{Gurovich et al.}{2010}]{gurovich10} 
        Gurovich, S., Freeman, K. C., Jerjen, H., Staveley-Smith, L.,
        \& Puerari, I. 2010, AJ, 140, 663

\bibitem[\protect\citeauthoryear{Gurovich et al.}{2004}]{gurovich04} 
        Gurovich, S., McGaugh, S. S., Freeman, K. C., Jerjen, H., Staveley-Smith, L.,
        \& De Blok, W. J. G. 2004, PASA, 21, 412

\bibitem[\protect\citeauthoryear{Haynes et al.}{2000}]{haynes00} 
        Haynes, M. P., Jore, K. P., Barrett, E. A., Broeils, A. H.,
        \& Murray, B. M. 2000, AJ, 120, 703

\bibitem[\protect\citeauthoryear{Ho}{2007}]{ho07} 
        Ho, L. C. 2007, ApJ, 668, 94

\bibitem[\protect\citeauthoryear{Hogg et al.}{2001}]{hogg01} 
        Hogg, D. E., Roberts, M. S., Bregman, J. N., \& Haynes, M. P. 2001,
        AJ, 121, 1336

\bibitem[\protect\citeauthoryear{Holmberg}{1958}]{holmberg58} 
        Holmberg, E. 1958, Lund Medd. Astron. Obs. Ser. II, 136, 1

\bibitem[\protect\citeauthoryear{Iodice et al.}{2003}]{iodice03}
	Iodice, E., Arnaboldi, M., Bournaud, F., Combes, F., Sparke,
        L. S., van Driel, W., \& Capaccioli, M. 2003, ApJ, 585, 730

\bibitem[\protect\citeauthoryear{Kannappan et al.}{2002}]{kannappan02}
     	Kannappan, S. J., Fabricant, D. G., \& Franx, M. 2002, AJ, 123, 2358

\bibitem[\protect\citeauthoryear{Kassin et al.}{2007}]{kassin07}
        Kassin, S. A. et al. 2007, ApJ, 660, L35

\bibitem[\protect\citeauthoryear{Kauffmann et al.}{2003}]{kauffmann03a}
        Kauffmann, G. et al. 2003, MNRAS, 341, 33

\bibitem[\protect\citeauthoryear{Kent et al.}{2008}]{kent08}  
	Kent, B. R. et al. 2008, AJ, 136, 713

\bibitem[\protect\citeauthoryear{Martin et al.}{2005}]{galex}
        Martin, D. C. et al. 2005, ApJ, 619, L1
 
\bibitem[\protect\citeauthoryear{Masters et al.}{2006}]{masters06} 
	Masters, K. L., Springob, C. M., Haynes, M. P., \& Giovanelli,
        R. 2006, ApJ, 653, 861

\bibitem[\protect\citeauthoryear{McGaugh}{2005}]{mcgaugh05}
        McGaugh, S. S. 2005, ApJ, 632, 859

\bibitem[\protect\citeauthoryear{McGaugh et al.}{2000}]{mcgaugh00}   
	McGaugh, S. S., Schombert, J. M., Bothun, G. D., \& de Blok, W. J. G.
        2000, ApJ, 533, L99

\bibitem[\protect\citeauthoryear{Meyer et al.}{2008}]{meyer08}
        Meyer, M. J., Zwaan, M. A., Webster, R. L., Schneider, S., \&
        Staveley-Smith, L. 2008, MNRAS, 391, 1712

\bibitem[\protect\citeauthoryear{Mo et al.}{1998}]{mmw98}
        Mo, H. J., Mao, S., \& White, S. D. M. 1998, MNRAS, 295, 319

\bibitem[\protect\citeauthoryear{Moran et al.}{2010}]{moran10}
        Moran, S. M. et al, 2010, ApJ, 720, 1126

\bibitem[\protect\citeauthoryear{Neistein et al.}{1999}]{eyal99}
        Neistein, E., Maoz, D., Rix, H.-W., \& Tonry, J. L. 1999, AJ, 117, 2666

\bibitem[\protect\citeauthoryear{Noordermeer \& Verheijen}{2007}]{noordermeer07}
        Noordermeer, E., \& Verheijen, M. A. W. 2007, MNRAS, 381, 1463

\bibitem[\protect\citeauthoryear{Paturel et al.}{2003a}]{paturel03a}  
	Paturel, G., Petit, C., Prugniel, P., Theureau, G., Rousseau,
        J., Brouty, M., Dubois, P., \& Cambr{\'e}sy, L. 2003a, A\&A, 412, 45

\bibitem[\protect\citeauthoryear{Paturel et al.}{2003b}]{paturel03b}  
	Paturel, G., Theureau, G., Bottinelli, L., Gouguenheim, L.,
        Coudreau-Durand, N., Hallet, N., \& Petit, C. 2003b, A\&A, 412, 57

\bibitem[\protect\citeauthoryear{Pizagno et al.}{2007}]{pizagno07}
        Pizagno, J. et al. 2007, AJ, 134, 945

\bibitem[\protect\citeauthoryear{Pizzella et al.}{2005}]{pizzella05}
        Pizzella, A., Corsini, E. M., Dalla Bont{\`a}, E., Sarzi, M.,
        Coccato, L., \& Bertola, F. 2005, ApJ, 631, 785

\bibitem[\protect\citeauthoryear{Saintonge et al.}{2008}]{saintonge08}
        Saintonge, A., Masters, K. L., Marinoni, C., Spekkens, K.,
        Giovanelli, R., \& Haynes, M. P. 2008, A\&A, 478, 57

\bibitem[\protect\citeauthoryear{Saintonge \& Spekkens}{2011}]{saintonge11}
        Saintonge, A., \& Spekkens, K. 2011, ApJ, 726, 77 (SS11)

\bibitem[\protect\citeauthoryear{Saintonge et al.}{2011}]{coldgass1}
        Saintonge, A. et al. 2011, MNRAS, 415, 32

\bibitem[\protect\citeauthoryear{Salim et al.}{2007}]{salim07}  
        Salim, S. et al. 2007, ApJS, 173, 267

\bibitem[\protect\citeauthoryear{Scannapieco et al.}{2009}]{scannapieco09}
        Scannapieco, C., White, S. D. M, Springel, V., \& Tissera,
        P. B. 2009, MNRAS, 396, 696

\bibitem[\protect\citeauthoryear{Scannapieco et al.}{2011}]{scannapieco11}
        Scannapieco, C., White, S. D. M, Springel, V., \& Tissera, P. B. 2011, arXiv:1105.0680

\bibitem[\protect\citeauthoryear{Springob et al.}{2005}]{s05}   
        Springob, C. M., Haynes, M. P., Giovanelli, R., \&  Kent,
        B. R. 2005, ApJS, 160, 149 (S05)



\bibitem[\protect\citeauthoryear{Tully \& Fisher}{1977}]{tf77}
        Tully, R. B. \& Fisher, J. R. 1977, A\&A, 54, 661

\bibitem[\protect\citeauthoryear{Tully et al.}{2009}]{tully09} 
        Tully, R. B., Rizzi, L., Shaya, E. J., Courtois, H. M.,
        Makarov, D. I., \& Jacobs, B. A. 2009, AJ, 138, 323

\bibitem[\protect\citeauthoryear{Wang et al.}{2011}]{wang11}
        Wang, J. et al. 2011, MNRAS, 412, 1081

\bibitem[\protect\citeauthoryear{Weinmann et al.}{2009}]{weinmann09}
        Weinmann, S. M., Kauffmann, G., van der Bosch, F. C., Pasquali,
        A., McIntosh, D. H., Mo, H., Yang, X., \& Guo, Y. 2009, MNRAS, 394, 1213

\bibitem[\protect\citeauthoryear{Williams et al.}{2010}]{williams10}
	Williams, M. J., Bureau, M., \& Cappellari, M. 2010, MNRAS, 409, 1330

\bibitem[\protect\citeauthoryear{York et al.}{2000}]{sdss}  
        York, D. G., et al. 2000, AJ, 120, 1579

\bibitem[\protect\citeauthoryear{Zaritsky et al.}{2008}]{zaritsky08} 
        Zaritsky, D., Zabludoff, A. I., \& Gonzalez, A. H. 2008, ApJ, 682, 68


\end{thebibliography}
\end{document}